\documentclass{amsart}[10pt]

\usepackage{amscd,amssymb,latexsym,url,verbatim,graphicx,color}
\usepackage{tikz,tikz-cd}
\usetikzlibrary{matrix, arrows, decorations.pathmorphing}

\usepackage{cases,amsmath}

\usepackage{txfonts}

\usepackage{tikz,tikz-cd}

\usepackage[dvips]{epsfig}

\title[Flip graphs and Weak Symmetry Breaking]
{Structure theory of flip graphs with applications to Weak Symmetry Breaking}

\author{Dmitry N. Kozlov}

\address{Department of Mathematics, University of Bremen, 28334
  Bremen, Federal Republic of Germany}

\email{dfk@math.uni-bremen.de}

\keywords{distributed computing, combinatorial algebraic topology,
  immediate snapshot, protocol complexes, Weak Symmetry Breaking,
  chromatic subdivision} 
\newtheorem{theorem}{Theorem}[section]
\newtheorem{df}[theorem]{Definition}
\newtheorem{thm}[theorem]{Theorem} 

\newtheorem{prop}[theorem]{Proposition}
\newtheorem{lm}[theorem]{Lemma} 
\newtheorem{crl}[theorem]{Corollary}

\newtheorem{rem}[theorem]{Remark}
\newtheorem{dcc}[theorem]{Distributed Computing Context}
\newtheorem{simp}[theorem]{Simplicial interpretation}
 
\newcommand{\nin}{\noindent}
\newcommand{\pr}{\nin{\bf Proof.} }

\newcommand{\cb}{{\mathcal B}}
\newcommand{\cf}{{\mathcal F}}

\newcommand{\cm}{{\mathcal M}}
\newcommand{\cn}{{\mathcal N}}
\newcommand{\cp}{{\mathcal P}}

\newcommand{\da}{\Delta}
\newcommand{\dar}{\downarrow}
\newcommand{\dl}{\textrm{dl}}

\newcommand{\mych}{\chi}
\newcommand{\ra}{\rightarrow}

\newcommand{\sm}{\setminus}
\newcommand{\ti}{\tilde}
\newcommand{\wti}{\widetilde}

\newcommand{\ab}{\allowbreak}

\newcommand{\s}{\,|\,}
\newcommand{\ds}{\,\,\|\,\,}
\newcommand{\pre}{\textrm{Pr}}

\newcommand{\fatb}{{\bf b}}
\newcommand{\fatx}{{\bf x}}

\newcommand{\gn}{\Gamma_n}

\newcommand{\flip}{\cf}
\newcommand{\msb}{{\textrm{sb}\,}}
\newcommand{\carr}{\textrm{carrier}\,}
\newcommand{\supp}{\carr}
\newcommand{\col}{\textrm{color}\,}
\newcommand{\id}{\col}
\newcommand{\parent}{\textrm{parent}\,}
\newcommand{\dist}{\textrm{dist}\,}
\newcommand{\defa}{\textrm{def}}

\newcommand{\swap}{{\tt swap}}
\newcommand{\up}{{\tt up}}
\newcommand{\specup}{{\tt specup}}
\newcommand{\dd}{\textrm{ds}}

\newcommand{\suf}[1]{\langle #1 \rangle}
\newcommand{\suff}{\suf}

\numberwithin{equation}{section}
\numberwithin{figure}{section}
\numberwithin{table}{section}

\begin{document}

\begin{abstract}
This paper is devoted to advancing the theoretical understanding of
the iterated immediate snapshot (IIS) complexity of the Weak Symmetry
Breaking task (WSB). Our rather unexpected main theorem states that
there exist infinitely many values of $n$, such that WSB for
$n$~processes is solvable by a~certain explicitly constructed
$3$-round IIS protocol. In particular, the minimal number of rounds,
which an~IIS protocol needs in order to solve the WSB task, does not go
to infinity, when the number of processes goes to infinity.
Our methods can also be used to generate such values of~$n$.

We phrase our proofs in combinatorial language, while avoiding using
topology. To this end, we study a~certain class of graphs, which we
call flip graphs. These graphs encode adjacency structure in certain
subcomplexes of iterated standard chromatic subdivisions of a~simplex.
While keeping the geometric background in mind for an additional
intuition, we develop the structure theory of matchings in flip graphs
in a~purely combinatorial way. Our bound for the IIS complexity is
then a~corollary of this general theory.

As an afterthought of our result, we suggest to change the overall
paradigm. Specifically, we think, that the bounds on the IIS
complexity of solving WSB for $n$ processes should be formulated in
terms of the size of the solutions of the associated Diophantine
equation, rather than in terms of the value $n$ itself.
\end{abstract}

\maketitle


\tableofcontents



\section{Introduction}

\subsection{Solvability of Weak Symmetry Breaking} The Theoretical 
Distributed Computing revolves around studying solvability and
complexity of the so-called {\it distributed tasks.} Roughly speaking
these are specifications of sets of required outputs for all legal
inputs.  One of the classical and central tasks is the so-called {\it
  Hard $M$-Renaming}. For this task, $n$ processes with unique names
in a~large name space of size~$N$ must cooperate in a~wait-free manner
to choose unique names from a~typically much smaller name space of
size~$M$. 

In order to talk about solvability and complexity of various tasks,
one needs to specify the computational model. A~standard one, called
{\it iterated immediate snapshot}, has been in the center of attention
of many papers, including this one. In this model the processes use
two atomic operations being performed on shared memory. These
operations are: {\it write} into the register assigned to that
process, and {\it snapshot read}, which reads entire memory in one
atomic step. Furthermore, it is assumed that the executions are
well-structured in the sense that they must satisfy the two following
conditions. First, it is only allowed that at each time a~group of
processes gets active, these processes perform a~write operation
together, and then they perform a~snapshot read operation together; no
other interleaving in time of the write and read operations is
permitted.  Such executions are called {\it immediate snapshot}
executions. Second, each execution can be broken up in rounds, where
in every round each non-faulty process gets activated precisely once,
alternatively, this can be phrased as each process using fresh memory
every time its gets activated.

Historically, the main focus, when studying the Hard $M$-Renaming in
the iterated immediate snapshot model, has been on finding the lower
bounds for~$M$. About 20 years ago it was shown that $M\geq
2n-1$. Inconveniently, the proof only worked for the case when $n$ is
a~prime power. Since no other classical problem in Distributed
Computing depends on the number-theoretic properties of~$n$, one was
tempted to believe that the appearance of the prime power condition
was an artefact of the method of the proof, rather than that of the
underlying question. Most surprisingly, the exact opposite was shown
to be true. A~long and complex construction was proposed to
demonstrate that algorithms exist for $M = 2n$, when $n$ is not
a~prime power. Unfortunately, it was difficult to calculate the
communication round complexity using that construction.

A~further task, the so-called {\it Weak Symmetry Breaking} task (WSB)
for $n$ processes, is an~inputless task with possible outputs $0$ and
$1$. A~distri\-bu\-ted protocol is said to solve the WSB if in any
execution without failed processes, there exists at least one process
which has value $0$ as well as at least one process which has
value~$1$. WSB for $n$ processes is equivalent to the Hard
$(2n-2)$-Renaming task, providing one of the mains reasons to study
its solvability and its complexity.

In the classical setting, the processes trying to solve WSB know their
id's, and are allowed to compare them. It is however not allowed that
any other information about id's is used. The protocols with this
property are called {\it comparison-based}.\footnote{Alternative
  terminology {\it rank-symmetric} is also used in the literature.} In
practice this means that behavior of each process only depends on the
relative position of its id among the id's of the processes it
witnesses and not on its actual numerical value. In this way, trivial,
uninteresting solutions can be avoided. As a~special case, we note
that each process must output the same value in case he does not
witness other processes at all.

One of the reasons the Iterated Immediate Snapshot model is used
extensively in Distributed Computing, which is also its major
advantage, is that the protocol complexes have a comparatively simple
simplicial structure, and are amenable to mathematical
analysis. Specifically, the existence of a~distributed protocol
solving WSB in $r$ rounds is equivalent to the existence of a certain
$0/1$-labeling of the vertices of the $r$th iterated standard
chromatic subdivision of an~$(n-1)$-simplex.

\subsection{Previous work}
The Iterated Immediate Snapshot model is due to Gafni\&Borowsky,
see~\cite{BG1,BG2}; in~\cite{HKR} this model goes under the name {\it
  layered immediate snapshot}. Several groups of researchers have
studied the solvability of the WSB by means of comparison-based IIS
protocols. Due primarily to the work of Herlihy\&Shavit, \cite{HS}, as
well as Casta\~neda\&Rajsbaum, \cite{CR0,CR1,CR2}, it is known that
the WSB is solvable if and only if the number of processes is not
a~prime power, see also \cite{AP} for a~counting-based argument for
the impossibility part. This makes $n=6$ the smallest number of
processes for which this task is solvable.

The combinatorial structures arising in related questions on
subdivisions of simplex paths have been studied in \cite{ACHP,paths}. 
The specific case $n=6$ has been studied in \cite{ACHP}, who has proved 
the existence of the distributed protocol which solves the WSB task in 
17 rounds. This bound was recently improved to 3 rounds in~\cite{wsb6}, 
where also an explicit protocol was given.

We recommend~\cite{AW} as a general reference for Theoretical
Distributed Computing, and \cite{book} as a general reference for
combinatorial topology. Furthermore, our book~\cite{HKR} contains
all the standard terminology, which we are using here.

A~broader framework of symmetry breaking tasks can be found
in~\cite{IRR}. The topological description for the IIS model can be
found in \cite{HKR,HS}; in addition, topological descriptions of
several other computational models have also been studied,
see~\cite{BR,k2,k1,view}.

\subsection{Our results}

Our main result states that, surprisingly, there is an infinite set of
numbers of processes for which WSB can be solved in $3$ rounds in the
comparison-based IIS model. Specifically, we prove that this is the
case when the number of processes is divisible by~$6$. There is
$O(n^2)$ overhead cost to translate an IIS protocol to an IS protocol,
so the resulting IS complexity is $O(n^2)$, which is of course less
surprising.

Let $\msb(n)$ denote the minimal number of rounds which is needed for
an IIS protocol to solve WSB for $n$ processes, then our main theorem
can succinctly be stated as follows.

\begin{thm}\label{thm:6n}
For all $t\geq 1$, we have $\msb(6t)\leq 3$.
\end{thm}

Our proof is based on combinatorial analysis of certain matchings in
the so-called flip graphs, and strictly speaking does not need any
topology.

\section{Informal sketch of the proof}

\subsection{The situation prior to this work.}
It has long been understood, that there is a~1-to-1 correspondence
between IIS protocols solving WSB on one hand and binary assignments
$\lambda$ to the vertices of the iterated chromatic subdivision of
a~simplex, on the other hand, where these assignments must satisfy
certain technical boundary conditions, and have no monochromatic
top-dimensional simplices, see, e.g., \cite{HKR}. Furthermore,
Herlihy$\,$\&$\,$Shavit, see \cite{HS}, found an~obstruction to the existence
of such an assignment in case the number of vertices of that simplex
(i.e., the number of processes) is a~prime power. This obstruction is
a~number which only depends on the values of $\lambda$ on the boundary
of the subdivided simplex, and which must be $0$ if there are no
monochromatic maximal simplices.

Thus the construction of IIS protocols solving WSB has been reduced to
finding such $\lambda$, where the number of IIS rounds is equal to the
number of iterations of the standard chromatic subdivision.  The
construction of $\lambda$, when the number is not a prime power, was
then done by Casta\~neda$\,$\&$\,$Rajsbaum, see~\cite{CR2}, using the
following method. First, boundary values are assigned, making sure
that this obstruction value is~$0$. After that the rest of the values
are assigned, taking some care that only few monochromatic maximal
simplices appear in the process. This is followed by a~sophisticated
and costly reduction procedure, during which the monochromatic
simplices are connected by paths, and eventually eliminated.  This
elimination procedure is notoriously hard to control, leading to
exponential bounds.

\subsection{The main ideas of our approach}
The idea which we introduce in this paper is radically different.
Just as Casta\~neda$\,$\&$\,$Rajsbaum we produce a boundary labeling
making sure the invariant is $0$. However after that our approaches
diverge in a~crucial way. We assign value $0$ to {\bf all} internal
vertices. This is quite counter-intuitive, as we are trying to get rid
of the monochromatic simplices in the long run, while such an
assignment on the contrary will produce an~enormous amount of
them. However, the following key observation comes to our rescue: if
we can match the monochromatic simplices with each other so that any
pair of matched simplices shares a boundary simplex of one dimension
lower, then we can eliminate them all in one go using one more round.

Next, we make a bridge to combinatorics. We have a graph, whose
vertices are all the monochromatic maximal simplices, connected by an
edge if they share a boundary simplex of one dimension lower; we shall
call such graphs {\it flip} graphs. What we are looking for is a {\it
 perfect matching} on graphs of this type. This is a~simple reduction, 
but it is very fruitful, since the matching theory on graphs is a~very 
well-developed subject and we find ourselves having many tools at our 
disposal. A~classical method to enlarge existing matchings is that of 
{\it augmenting paths}. The idea is elementary but effective: connect 
the unmatched (also called critical) vertices by a~path $p$, such that 
all other vertices on the path are matched by the edges along $p$, and 
then make all the non-matching edges of $p$ matching and vice versa. 
This trick will keep all the internal vertices of $p$ matched, 
while also making end vertices matched. In particular, if we
have a matching, and we succeeded to connect critical vertices in
pairs by non-intersection augmenting paths, then applying this trick
to all these paths simultaneously, we will end up with a~perfect
matching.

\subsection{The blueprint of the proof}
This set of ideas leads to the following blueprint for constructing
the 3-round IIS protocol to solve WSB for $n$ processes:
\begin{enumerate}
\item[{\bf Step 1.}] Find a good boundary assignment for the second
  standard chromatic subdivision of the simplex with $n$ vertices,
  making sure the Herlihy-Shavit obstruction vanishes. Assign value
  $0$ to {\it all} internal vertices.
\item[\bf Step 2.] Decompose the resulting flip graph of monochromatic
simplices into pieces corresponding to the maximal simplices of the first
chromatic subdivision. Describe an initial matching on each of these
pieces, and combine them to a~total matching.
\item[\bf Step 3.] Construct a~system of non-intersecting augmenting
  paths with respect to that total matching. Changing our initial  
	matching along these paths produces the desired perfect matching.
\item[\bf Step 4.] Eliminate all monochromatic maximal simplices in
  one go, producing a~binary assignment for the third standard
  chromatic subdivision of the simplex with $n$ vertices, which now
  has no maximal monochromatic simplices.
\end{enumerate}

This is a general scheme, and if the technical details work out, it
can be used for various values of $n$ and also for various numbers of
rounds.  In this paper we restrict ourselves to the values
$n=6,12,18,\dots$, mainly because this is the case in which we can
provide complete rigorous details. The techniques of this paper can
further be extended to discover other values of $n$ for which WSB can
be solved in $3$ rounds, see \cite{bid}. In that paper, progress has
been made using techniques of Sperner theory, see~\cite{An},
specifically a variation of local LYM inequality, to cover values
$n=15,20,21,\dots$, see Theorem~\ref{conj:prim}.

To start with, for $n=6t$, there is a quite special arithmetic
identity~\eqref{eq:6t}, which has a~stronger set-theoretic version,
see the proof of Theorem~\ref{thm:6n}, stating the existence of
a~certain bijection~$\Phi$. We produce a~quite special labeling of the
vertices on the boundary of $\mych^2(\da^{n-1})$, and we put the label
$0$ on all the internal vertices of~$\mych^2(\da^{n-1})$. Since any
top-dimensional simplex has at least one internal vertex, we will have
no $1$-monochromatic simplices at that point. This corresponds to the
step 1 above.

We then proceed with steps 2 and 3, which are at the technical core of
our proof. We start with a rough approximation to the matching which
we want to get at the end. In this approximation, called the {\it
  standard matching}, most of the simplices will get matched. There
will be a small number of critical simplices left, concentrated around
barycenters of certain boundary simplices. We then find a~system of
augmenting paths which connect all the critical simplices in
pairs. Our idea of how to find these paths is to use the system of
non-intersection paths in $\gn$ which we construct along the bijection
$\Phi$, like a~''system of tunnels'' between areas of
$\mych^2(\da^{n-1})$ which contain the monochromatic simplices. Within
each such tunnel we use our analysis of the combinatorial structure of
the flip graphs, namely certain properties, which we call {\it
  conductivity} of these graphs, to connect the critical simplices by
augmenting paths, see Figure~\ref{fig:pext}. This yields the desired
result, allowing us to produce a perfect matching on the set of
monochromatic simplices. Step 4 is an easy and standard step which has
been used before, we do not make any original contribution there.

\subsection{Note on the language we use to formulate our argument}
As mentioned above, it is by now a~classical knowledge that executions of 
a~distributed protocol in IIS model can be encoded using the simplicial 
structure of the standard chromatic subdivision and its iterations.
Consequently, various tools of topology have been used in the past to gain
information on the distributed computing tasks. In contrast, our argument does
not need any implicit or {\it non-constructive} topological results, such as,
for instance, existance of fixed points. All that is required is the incidence structure
if the underlying subdivisions and various labeling techniques for the vertices.

To underline this fact, we shall mostly omit any mentioning of simplicial
structures, and formulate everything using {\it only} the language of graph theory.
It is certainly very helpful for the intuition to keep the simplicial
picture in mind, and we invite the reader to do so when going through
the text. Also, many of our illustrations refer to the simplicial picture.
However, we feel it is of value to have our proof phrased exclusively in
terms of combinatorics of graphs.

\section{Basic concepts}

\subsection{Graph theory concepts} $\,$

\nin We start by recalling some graph terminology, which we need
throughout the paper. For a~graph $G$ we let $V(G)$ denote the set of
its vertices and we let $E(G)$ denote the set of its edges. Two
different edges are called {\it adjacent} if they share a~vertex.

\begin{df}
An {\bf edge coloring} of a~graph $G$ with colors from a~set $C$ is an
assignment $c:E(G)\to C$, such that adjacent edges get different
colors.
\end{df}

Assume $G$ is a graph and $A$ is a~subset of $V(G)$. We say that the
graph $H$ is the subgraph of $G$ {\it induced by} $A$, if the set of
vertices of $H$ is~$A$, and two vertices of $H$ are connected by an
edge in $H$ if and only if they are connected by an edge in~$G$.

\begin{df}
A~{\bf matching} on a graph $G$ is a~set of edges, such that no two of
these edges are adjacent. The vertices of these edges are said to be
matched, while the rest of the vertices are called~{\bf critical}.
\end{df}
 
\nin To underline that not all vertices are matched we often
say {\it partial} matching. 

\begin{df}\label{df:perfm}
The matching is called {\bf perfect} if all vertices are matched, and
it is called {\bf near-perfect} if it has exactly one critical vertex.
\end{df}

\subsection{Set theory concepts} $\,$

\nin For all natural numbers $n$, we let $[n]$ denote the set
$\{1,\dots,n\}$.  Furthermore, throughout the paper we shall skip
curly brackets when the set consists of a~single element.  When we say
a {\it tuple} we mean any ordered sequence. An {\it order} $R$ on
a~set $S$ is any tuple $R=(x_1,\dots,x_d)$, satisfying
$S=\{x_1,\dots,x_d\}$.

\begin{df}\label{df:norm}
Assume $A$ is an arbitrary nonempty finite set of natural numbers, and
let $k$ denote the cardinality of $A$. Then there exists a~unique
order-preserving bijection $\varphi:A\ra[k]$. We call $\varphi$ the
{\bf normalizer} of~$A$.
\end{df}

Note, that if $A$ and $B$ are equicardinal nonempty finite sets of
natural numbers, $\varphi$ is a~normalizer of $A$, and $\gamma:A\ra B$
is the unique order-preserving bijection between $A$ and $B$, then
$\varphi\circ\gamma^{-1}$ is the normalizer of~$B$. On the other hand,
if $\psi:B\ra[k]$ is a~normalizer of $B$, then $\psi\circ\gamma$ is
a~normalizer of~$A$.

When talking about normalizers in the rest of the paper, we shall
typically include the information on the set cardinality in
normalizers definition. In other words, when we say $\varphi:A\ra[k]$
{\it is a normalizer of} $A$, we mean {\it let} $k$ {\it denote the
  cardinality of} $A$ {\it and let the map $\varphi:A\ra[k]$ denote
  the unique order-preserving bijection.}

\subsection{$S$-tuples} $\,$

\begin{df}
\label{df:osp}
For any finite set $S$, an {\bf $S$-tuple} is a~tuple
$(A_1,\dots,A_t)$ of disjoint non-empty subsets of $S$. We call $t$
the {\bf length} of this $S$-tuple. We shall use the short-hand
notation $A_1\s\dots \s A_t$.  An $S$-tuple $A_1\s\dots\s A_t$ is
called {\bf full} if $A_1\cup\dots\cup A_t=S$. For a~full $S$-tuple
$\sigma=A_1\s\dots\s A_t$, we let $V(\sigma)$ denote the set
$A_1\cup\dots\cup A_{t-1}=S\sm A_t$.
\end{df}

\nin Note, that in~\cite{wsb6} full $S$-tuples were called {\it
  ordered set partitions of the set~$S$}.

Clearly, if $T\supseteq S$, then any $S$-tuple can be interpreted as
a~$T$-tuple as well. Furthermore, if $A=A_1\s\dots\s A_k$ is an
$S$-tuple and $B$ is another $S$-tuple, we say that $A$ is a {\it
  truncation of~$B$}, if $B$ has the form $A_1\s\dots\s A_k\s
A_{k+1}\s\dots\s A_t$, for some $t\geq k$.

\begin{dcc}
Full $[n]$-tuples are mathematical objects which encode possible 
executions of the standard one round protocol for $n$ processes 
in the immediate snapshot model.
\end{dcc}

\begin{df}
\label{df:posp}
Given a set $S$, and two $S$-tuples of the same length
$\sigma=A_1\s\dots\s A_t$ and $\tau=B_1\s\dots\s B_t$, we call the
ordered pair $(\sigma,\tau)$ a~{\bf coherent pair of $S$-tuples}, if we
have the set inclusion $B_i\subseteq A_i$, for all $i=1,\dots,t$.
\end{df}

We find it convenient to view a~coherent pair of $S$-tuples as a~$2\times
t$ table of sets
\[(\sigma,\tau)=
\begin{array}{|c|c|c|c|c|}
\hline
A_1 & \dots & A_t \\ \hline
B_1 & \dots & B_t \\ 
\hline
\end{array}.
\]

An arbitrary $S$-tuple $A_1\s\dots\s A_t$ can alternatively be viewed
as a~coherent pair of $S$-tuples $(A_1\s\dots\s A_t,A_1\s\dots\s A_t)$,
and we shall use the two interchangeably. We also recall the following
terminology from \cite{wsb6}: for an~$S$-tuple $\sigma=A_1\s\dots\s
A_t$ we set its {\it carrier set} to be
$\carr(\sigma):=A_1\cup\dots\cup A_t$; for a~coherent pair of $S$-tuples
$(\sigma,\tau)$, we set $\carr(\sigma,\tau):=\carr(\sigma)$, and we
set $\col(\sigma,\tau):=\carr(\tau)$; the latter is called the {\it
  color set} of $(\sigma,\tau)$.

\begin{dcc}\label{dc1}
The terminology of coherent pairs of $S$-sets comes from the need to have combinatorial
language to describe partial views on the executions of distributed protocols.
It is related to the previous work of the author, see \cite{subd,k2,k1,view}.
The sets $B_1,\dots,B_t$ contain the id's of the active processes whose view
on the execution is available to us. Accordingly, the color set contains
id's of all active processes. The sets $A_1,\dots,A_t$ contain the id's of all
processes which are seen by the active ones. These include the active processes,
as any process sees itself, but it may include more than that.  
Accordingly, the carrier set contains id's of a~possibly larger set of processes, 
which are seen by the active ones.
\end{dcc}

\begin{df} \label{df:restr}
Assume that we are given a~coherent pair of $S$-tuples, say
$(\sigma,\tau)=(A_1\s\dots\s A_t,B_1\s\dots\s B_t)$, together with
a~non-empty proper subset $T\subset\col(\sigma,\tau)$. We define a~new
coherent pair of $S$-tuples $(\tilde\sigma,\tilde\tau)$, which we call
a~{\bf restriction} of $(\sigma,\tau)$ to $T$.  To do this, we first
decompose $T=T_1\cup\dots\cup T_d$ as a~disjoint union of non-empty
subsets such that $T_1\subseteq B_{i_1}$, $\dots$, $T_d\subseteq
B_{i_d}$, for some $1\leq i_1<\dots<i_d\leq t$. Then, we set $\tilde
A_1:=A_1\cup\dots\cup A_{i_1}$, $\tilde A_2=A_{i_1+1}\cup\dots\cup
A_{i_2}$, $\dots$, $\tilde A_d=A_{i_{d-1}+1}\cup\dots\cup
A_{i_d}$. Finally, we set $\tilde\sigma:=\tilde A_1\s\dots\s\tilde
A_d$, and $\tilde\tau:=T_1\s\dots\s T_d$. We denote this new coherent
pair of $S$-tuples by $(\sigma,\tau)\dar T$.
\end{df}
\nin A~restriction of $(\sigma,\tau)$ to $T$ is always uniquely defined by
the above. 

Note, that since an~arbitrary $S$-tuple can be viewed as a~coherent
pair of $S$-tuples, we are able to talk about restrictions of
$S$-tuples. However, the set of $S$-tuples, unlike the set of coherent
pairs of $S$-tuples, is not closed under taking restrictions, and the
result of restricting an~$S$-tuple will be a~coherent pair of
$S$-tuples.

As an example, let $S=[6]$,
$\sigma=\{1,2\}\s 3\s 4\s 5\s 6$, and $T=\{1,2,3,5\}$, then
\[\sigma\dar T=
\begin{array}{|c|c|c|c|c|}
\hline
1,2 & 3 & 4,5 \\ \hline
1,2 & 3 & 5 \\ 
\hline
\end{array}.
\]
Furthermore, taking $\widetilde T=\{1,5\}$, we get
\[\sigma\dar\widetilde T=(\sigma\dar T)\dar\widetilde T=
\begin{array}{|c|c|c|c|}
\hline
1,2 & 3,4,5 \\ \hline
1 &  5 \\ 
\hline
\end{array}.
\]

Dually, we introduce an~operation of {\it deletion} by setting
$\dl((\sigma,\tau),T):=(\sigma,\tau)\dar(S\sm T)$, for an~arbitrary
coherent pair of $S$-tuples $(\sigma,\tau)$, and an~arbitrary
non-empty proper subset $T\subset S$. In this case we say that the
coherent pair of $S$-tuples $\dl((\sigma,\tau),T)$ is obtained from
$(\sigma,\tau)$ by {\it deleting}~$T$.


\begin{dcc}
The restriction operation described in Definition~\ref{df:restr} may sound
complicated. However, it faithfully describes what happens to the combinatorial
labels of partial executions, when the set of active processe is reduces, 
cf.\ Distributed Computing Context~\ref{dc1}. Some of the sets of processes 
may be merged, which happens when those processes which distinguished between
the two groups are no longer active. Other processes may stop to be seen
altogether; their id's are then deleted from the description. The need to 
phraze in combinatorial language what happens to the partial execution description,
led the author to formulate the Definition~\ref{df:restr}. 
\end{dcc}

\section{Flip graphs} \label{sect:3}

\subsection{The graphs $\gn$}\label{ssect:gn} $\,$

\nin When $\sigma=A_1\s\dots\s A_t$ is a~full $[n]$-tuple, we will use
the following short-hand notation:
\[F(\sigma):=
\begin{cases}
[n]\sm A_t, & \text{ if } |A_t|=1; \\
[n], & \text{ otherwise. }
\end{cases}\]

\nin
The set $F(\sigma)$ is called the {\it flippable set of $\sigma$}.

\begin{df}\label{df:flip}
Assume, we are given a~full $[n]$-tuple $\sigma=A_1\s\dots\s A_t$, and
$x\in F(\sigma)$. Let $k$ be the index, such that $x\in A_k$. We let
$\flip(\sigma,x)$ denote the full $[n]$-tuple obtained by using the
following rule.
\begin{enumerate}
\item[\text{\bf Case 1.}] If $|A_k|\geq 2$, then set 
\[\flip(\sigma,x):=A_1\s\dots\s A_{k-1}\s x \s A_k\sm x\s A_{k+1}\s\dots\s A_t.\]
\item[\text{\bf Case 2.}] If $|A_k|=1$ (that is $A_k= x $), then we
  must have $k<t$. We set
\[\flip(\sigma,x):=A_1\s\dots\s A_{k-1}\s x \cup A_{k+1}\s A_{k+2}\s\dots\s A_t.\]
\end{enumerate}
\end{df}

Due to background geometric intuition, we think of the process of
moving from a~full $[n]$-tuple $\sigma$ to the full $[n]$-tuple
$\tau=\flip(\sigma,x)$ as a~{\it flip}. If the first case of
Definition~\ref{df:flip} is applicable, we say that $\tau$ is obtained
from $\sigma$ by {\it splitting off} the element $x$, else we say that
$\tau$ is obtained from $\sigma$ by {\it merging in} the element~$x$.

The operation $\flip(-,x)$ behaves as a~flip operation is expected to
behave. Namely, for any $[n]$-tuple $\sigma$ and $x\in F(\sigma)$, we
have $x\in F(\flip(\sigma,x))$, and importantly
\begin{equation}\label{eq:flfl}
\flip(\flip(\sigma,x),x)=\sigma.
\end{equation}

\begin{prop}
Assume $\sigma$ is a~full $[n]$-tuple and $x\in F(\sigma)$, then we
have
\begin{equation}\label{eq:vf}
V(\cf(\sigma,x))\cap([n]\sm x)=V(\sigma)\cap([n]\sm x),
\end{equation}
where $V(-)$ is as in Definition~\ref{df:osp}. In other words, barring
$x$, the set of elements, which are not contained in the last set
of~$\sigma$, does not change during the flip.
\end{prop}
\pr Assume $\sigma=A_1\s\dots\s A_t$, and $\cf(\sigma,x)=B_1\s\dots\s
B_m$.  We have one of the 3 cases: $B_m=A_t$, $B_m=A_t\sm\{x\}$, or
$B_m=A_t\cup\{x\}$.  Either way, we have 
\[B_m\cap([n]\sm x)= A_t\cap([n]\sm x),\] 
so \eqref{eq:vf} follows.  \qed\vskip5pt

We now define the basic flip graphs.

\begin{df}\label{df:gn}
Let $n$ be any natural number.  We define a graph $\gn$ as follows. 
The vertices of $\gn$ are indexed by all full $[n]$-tuples. 
Two vertices $\sigma$ and $\tau$ are connected by an edge if and only 
if there exists $x\in F(\sigma)$, such that $\tau=\flip(\sigma,x)$.
\end{df}

We can color the edges of the graph $\gn$ by elements of $[n]$.  To do
this we simply assign color $x$ to the edge connecting full
$[n]$-tuple $\sigma$ with the full $[n]$-tuple $\flip(\sigma,x)$. It
follows from~\eqref{eq:flfl} that this assignment yields
a~well-defined {\it edge coloring} of the graph~$\gn$; meaning that we
will assign the same color to an~edge independently from which of its
endpoints we take as $\sigma$, and furthermore, any two edges which
share a~vertex will get different colors under this assignment.
Indeed, if two edges share a~vertex, then they must correspond to
flips with respect to different vertices, meaning that they also must
have different colors. Furthermore, consider an edge $(\sigma,\tau)$.
Take $x\in F(\sigma)$, such that $\tau=\cf(\sigma,x)$. By the
equality~\eqref{eq:flfl}, we have $\cf(\tau,x)=\sigma$, so the color
of the edge does not depend on the choice of the endpoint used to
define that color.

Combinatorially, the edges of $\gn$ are indexed by all coherent pairs
of $[n]$-tuples $(\sigma,\tau)=(A_1\s\dots\s A_t,\ab B_1\s\dots\s
B_t)$, satisfying the conditions: $\carr(\sigma,\tau)=[n]$, and
$|\col(\sigma,\tau)|=n-1$. In this case we have $B_1\cup\dots\cup
B_t=[n]\sm x$, where $x$ is the color of that edge.  Pick index $k$
such that $x\in A_k$. The vertices adjacent to that edge are
$A_1\s\dots\s A_{k-1}\s x\s A_k\sm x\s A_{k+1}\s\dots\s A_t$ and
$A_1\s\dots\s A_t$. On the other hand, given a~vertex
$\sigma=A_1\s\dots\s A_t$ of $\gn$, and $x\in F(\sigma)$, the edge
with color $x$ which is adjacent to $\sigma$ is indexed by
$\dl(\sigma,x)$.

\begin{figure}[hbt]

  \input{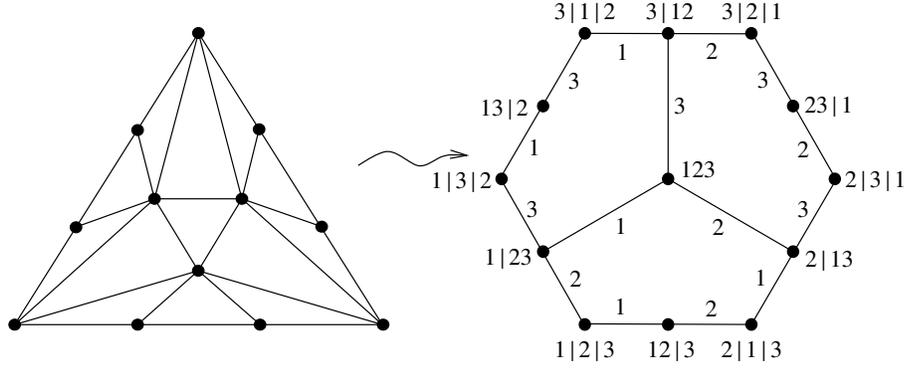}  

\caption{The standard chromatic subdivision and the flip graph.}
\label{fig:fig1}
\end{figure}

As an example, the $[3]$-tuples $\sigma_1=1\s 23$ and $\sigma_2=1\s
3\s 2$ index vertices of $\Gamma_3$. These vertices are connected by
an edge labeled with $3$ and indexed by the coherent pair of
$[3]$-tuples $(1|23,1|2)=\dl(\sigma_1,3)=\dl(\sigma_2,3)$. See
Figure~\ref{fig:fig1}.

\begin{dcc}
Two executions are connected by an edge if and only if there exists
exactly one process $x$ which has different views under these
executions; all other processes have the same view. The color of that
edge is~$x$. The two cases of Definition~\ref{df:flip} correspond to
the situation where process $x$ either stops seing processes which
executed ``at the same step'', or starts seeing processes, which
executed in the subsequent step. The views of other processes on $x$
or on each other are unaffected by that change.
\end{dcc}

\subsection{The graphs $\gn^2$}\label{ssect:gn2} $\,$

\nin As our next step, we consider ordered pairs of full $[n]$-tuples.
Given two full $[n]$-tuples $\sigma$ and $\tau$, we let
$\sigma\ds\tau$ denote the ordered pair $(\sigma,\tau)$. We shall also
combine this with our previous notations, so if $\sigma=A_1\s\dots\s
A_t$, and $\tau=B_1\s\dots\s B_q$, then $\sigma\ds\tau=A_1\s\dots\s
A_t\ds B_1\s\dots\s B_q$.

\begin{df}\label{df:flip2}
We define $F(\sigma\ds\tau):=F(\sigma)\cup F(\tau)$. 
Furthermore, for any $x\in F(\sigma\ds\tau)$ we define
\begin{equation}\label{eq:flip2}
\flip(\sigma\ds\tau,x):= 
\begin{cases}
\sigma\ds\flip(\tau,x),& \text{ if } x\in F(\tau);  \\
\flip(\sigma,x)\ds\tau,& \text{ if } x\notin F(\tau),\,\, x\in F(\sigma).
\end{cases}
\end{equation}
\end{df}

Note, that since $F(\sigma\ds\tau):=F(\sigma)\cup F(\tau)$, one of the
cases in equation~\eqref{eq:flip2} must occur.  We also remark that
$F(A_1\s\dots\s A_t\ds B_1\s\dots\s B_q)=[n]$, unless $A_t=B_q=x$, for
some $x\in[n]$, in which case we would have $F(A_1\s\dots\s A_t\ds
B_1\s\dots\s B_q)=[n]\sm x$.

\begin{df}
Let $n$ be any natural number. We define a graph $\gn^2$ as follows.
The vertices of $\gn^2$ are indexed by all ordered pairs of full
$[n]$-tuples.  Two vertices $\sigma_1\ds\tau_1$ and
$\sigma_2\ds\tau_2$ are connected by an edge if and only if there
exists $x\in F(\sigma_1\ds\tau_1)$, such that
$\sigma_2\ds\tau_2=\flip(\sigma_1\ds\tau_1,x)$.
\end{df}

\begin{figure}[hbt]

  \input{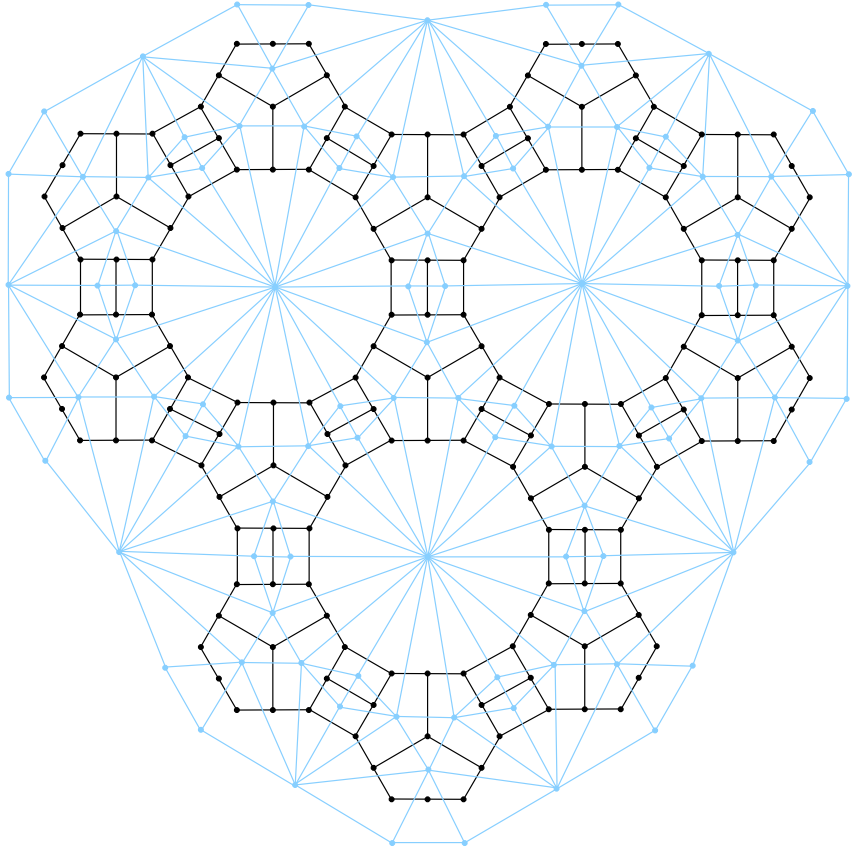}  

\caption{The flip graph $\Gamma_3^2$ shown in solid color. 
Lurking in the background is the second standard chromatic subdivision
of a~triangle.}
\label{fig:fig2}
\end{figure}

Figure~\ref{fig:fig2} shows the example $\Gamma_3^2$.

To describe the edges of $\gn^2$, we extend our $\ds$-notation and
write $(\sigma,\sigma')\ds(\tau,\tau')$, to denote an ordered pair of
coherent pairs of $[n]$-tuples, subject to an important additional
condition:
\[\col(\sigma,\sigma')=\carr(\tau,\tau').\] 
Now, the edges in $\gn^2$ are of two different types, corresponding to
the two cases of~\eqref{eq:flip2}:
\begin{itemize} 
\item either they are indexed by $\sigma\ds(\tau,\tau')$, where
  $\sigma$ is a~full $[n]$-tuple and $(\tau,\tau')$ indexes an edge
  in~$\gn$;
\item or they are indexed by $(\sigma,\sigma')\ds\tau$, where
  $(\sigma,\sigma')$ indexes an edge in $\gn$, and $\tau$ is a~full
  $\col(\sigma,\sigma')$-tuple.
\end{itemize}

As an example consider the following vertices of the graph
$\Gamma_3^2$: $v_1=1\s 23\ds 1\s 2\s 3$, $v_2=1\s 23\ds 12\s 3$, and
$v_3=1\s 3\s 2\ds 1\s 2\s 3$. The vertices $v_1$ and $v_2$ are
connected by an edge labeled by $1$ and indexed by $1\s 23\ds(12\s
3,2\s 3)$. The vertices $v_1$ and $v_3$ are connected by an edge
labeled by $3$ and indexed by $(1\s 23,1\s 2)\ds 1\s 2$.

We let $\gn^2(\sigma)$ denote the subgraph of $\gn^2$ induced by the
vertices of the form $\sigma\ds\tau$. Clearly, mapping $\sigma\ds\tau$
to $\tau$ gives an isomorphism between $\gn^2(\sigma)$ and~$\gn$.

\begin{dcc}
The vertices of $\gn^2$ correspond to $2$-round executions, and the
edges connect two executions where exactly one process changes it
view.  The intuition behind the two cases of Definition~\ref{df:flip2}
is as follows. If the process $x$ is seen by someone in the second
round, then what he saw in the first round is know to someone else, so
$x$ cannot change his first round view without affecting the others.
All the process $x$ can do, is to change its view of the second round
in exactly the same way as in the $1$-round model. This is the first
line of~\eqref{eq:flip2}. The second line of~\eqref{eq:flip2}
corresponds to the case when $x$ is the last one to act in the second
round, so nobody sees it. The execution we get, if $x$ changes its
view now, must come from $x$ changing its view of the first round
instead.
\end{dcc}

\newpage

\subsection{Higher flip graphs, support and subdivision maps} $\,$

\nin Since the graphs $\gn$ and $\gn^2$ are by far the main characters
of this paper, we have chosen to present them separately and in fine
detail.  However, the constructions from subsections~\ref{ssect:gn}
and~\ref{ssect:gn2} can easily be generalized to define the graphs
$\gn^d$, for arbitrary~$d\geq 1$. Though we will only need these
briefly for $d=3$, we include the general definitions for
completeness. The concepts in this subsection were previously
introduced in \cite[Section 3]{wsb6}.

Let us fix $d\geq 1$ and consider all $d$-tuples of full $[n]$-tuples,
$v=\sigma_1\ds\dots\ds\sigma_d$. Definition~\ref{df:flip} can be
generalized to such $d$-tuples as follows. Given $v$ as above we set
\[F(v):=F(\sigma_1)\cup\dots\cup F(\sigma_d).\] 
Effectively this means that $F(v)=[n]$, unless
$F(\sigma_1)=\dots=F(\sigma_d)=[n]\sm p$, for some $p$, in which
case we have $F(v)=[n]\sm p$.

\begin{df}\label{df:gflip}
Assume $v$ is a~$d$-tuple of $[n]$-tuples, and $x\in F(v)$. Let $k$ be
the maximal index such that $x\in F(\sigma_k)$, by the definition of
$F(v)$, such $k$ must exist. We define $\flip(v,x)$ to be the
following $d$-tuple of full $[n]$-tuples:
\begin{equation}\label{eq:dflip}
\flip(v,x):=(\sigma_1\ds\dots\ds\sigma_{k-1}\ds \flip(\sigma_k,x)\ds
\sigma_{k+1}\ds\dots\ds\sigma_n).
\end{equation}
\end{df}

Again, it is easy to see that for any $x\in F(v)$, we have the
identities $F(\flip(v,x))=F(v)$ and
\begin{equation}\label{eq:ffx}
\flip(\flip(v,x),x)=v.
\end{equation}

\begin{df}\label{df:gnd}
For an arbitrary $d\geq 1$ we define graph $\gn^d$ as follows.  The
vertices of $\gn^d$ are indexed by all $d$-tuples of full $[n]$-tuples.
Two vertices $v$ and $w$ are connected by an edge if and only if there
exists $x\in F(v)$, such that $w=\flip(v,x)$.
\end{df}

\nin In this context, the graph $\gn^1$ is the same as the graph
$\gn$. The edges of $\gn^d$ are indexed by all $d$-tuples of coherent
pairs of $[n]$-tuples, which for some $1\leq k\leq d$ have the special
form
$\sigma_1\ds\dots\ds\sigma_{k-1}\ds(\sigma_k,\sigma_k')\ds\sigma_{k+1}
\ds\dots\ds\sigma_d$,
where:
\begin{itemize}
\item $\sigma_1,\dots,\sigma_{k-1}$ are full $[n]$-tuples,
\item $(\sigma_k,\sigma_k')$ indexes an edge in $\gn$,
\item $\sigma_{k+1},\dots,\sigma_d$ are full
  $\col(\sigma_k,\sigma_k')$-tuples.
\end{itemize}

We call the graphs $\gn^d$ {\it higher flip graphs}. They are related
to each other by means of the so-called {\it support
  maps}. Specifically, given $c<d$, the support map $\supp_d^c$ goes
from the set of vertices of $\gn^d$ to the set of vertices $\gn^c$, it
takes the $d$-tuple $\sigma_1\ds\dots\ds\sigma_d$ to the $c$-tuple
$\sigma_1\ds\dots\ds\sigma_c$. If two vertices of $v$ and $w$ of
$\gn^d$ are connected by an edge, then either
$\supp_d^c(v)=\supp_d^c(w)$ or vertices $\supp_d^c(v)$ and
$\supp_d^c(w)$ are connected by an edge in~$\gn^c$.

Furthermore, assume we are given an~arbitrary vertex of $\gn^c$,
$v=\sigma_1\ds\dots\ds\sigma_c$, and an arbitrary number $d>c$. We
let $\gn^d(v)$ denote the subgraph of $\gn^d$ induced by all vertices
$w$, for which $\supp_d^c w=v$.  Clearly, we have a graph isomorphism
$\varphi:\gn^d(v)\simeq\gn^{d-c}$, given by
$\varphi(\sigma_1\ds\dots\ds\sigma_c\ds\tau_1\ds\dots\ds\tau_{d-c})=
\tau_1\ds\dots\ds\tau_{d-c}$.

\begin{dcc}
In the higher flip graph $\gn^d$ the vertices correspond to the
$d$-round executions. Again, two executions are connected by an edge
if and only if there exists exactly one process $x$ which has
different views under these executions; all other processes have the
same view. The color of that edge is~$x$. Our flip operation describes
precisely what happens to the label encoding the execution. Here, $k$
is the latest round in which someone has seen process~$x$.

Furthermore, a~vertex $v$ of $\gn^c$ corresponds to the initial $c$
rounds of an execution, the so-called prefix, and the graph $\gn^d(v)$
encodes all executions which start with that prefix~$v$.
\end{dcc}


\section{Standard matchings on graphs $\gn(\Omega,V)$} \label{sect:4}

\nin At this point we would like to issue a word of warning to the
reader.  While our initial mathematical concepts are closely related
to the Distributed Computing, this connection will weaken from this
point on. Once one has a~mathematical model, which is equivalent to
the questions in Distributed Computing which we would like to study,
we need to start exploring purely mathematical structures, in order to
be able to arrive at the resolution of our initial questions.

In particular, the concepts such as {\it forbidden sets, prefixes,
  standard matchings,} in this section, or {\it sets of patterns} and
other notions, in the subsequent sections, do not have a~distributed
computing interpretation, which is easy to grasp. Or perhaps the right
way to phrase this is that such an interpretation is yet to be
invented. We will however still provide distributed computing context
where possible, such as for example for the case of {\it nodes} in
Section~\ref{sect:6}.

\subsection{Forbidden sets and graphs $\gn(V)$}

\begin{df}\label{df:gnv}
Let $n$ be an~arbitrary natural number, and let $V$ be any subset of
$[n]$.  We let $\gn(V)$ denote the subgraph of $\gn$ induced by the
vertices indexed by those full $[n]$-tuples $A_1\s\dots\s A_t$, for
which $A_1\not\subseteq V$.
\end{df}

\begin{figure}[hbt]

  \input{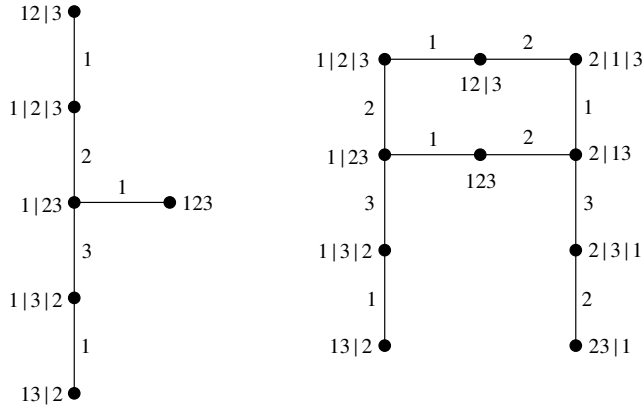}  

\caption{The graphs $\Gamma_3(\{2,3\})$ (left) and $\Gamma_3(\{3\})$
  (right).}
\label{fig:gnvex}
\end{figure}

We think of $V$ as a~''forbidden set'', in which case the condition in
Definition~\ref{df:gnv} simply says that the first block of the full
$[n]$-tuple must contain an~element which is {\it not}
forbidden. Clearly, when nothing is forbidden, we have no
restrictions, hence $\gn(\emptyset)=\gn$, and when everything is
forbidden, we have no full $[n]$-tuples satisfying that condition,
hence $\gn([n])=\emptyset$. Further examples are provided in
Figure~\ref{fig:gnvex}.

\subsection{Prefixes and graphs $\gn(\Omega,V)$}

\begin{df}
Let $V$ be any subset of $[n]$, and let $\sigma=A_1\s\dots\s A_t$ be
any full $[n]$-tuple. Let $t-1\geq k\geq 0$ denote the minimal number
for which $A_{k+1}\not\subseteq V$; if no such number exists, we set
$k:=t$. Then the $V$-tuple $A_1\s\dots\s A_k$ is called the {\bf
  $V$-prefix of $\sigma$}, and is denoted by $\pre_V(\sigma)$.
\end{df}

Note, that we allow the $V$-prefix of $\sigma$ to be empty, which is
the case if $k=0$, or equivalently $A_1\not\subseteq V$, meaning that
$\sigma$ is a~vertex of $\gn(V)$.

\begin{df}
Let $n$ be an arbitrary natural number, let $V$ be a~subset of $[n]$,
and let $\Omega$ be a~family of $V$-tuples. The graph $\gn(\Omega,V)$
is the subgraph of $\gn$ induced by all vertices $\sigma$, such that
$\pre_V(\sigma)\in\Omega$.
\end{df}

\begin{figure}[hbt]

  \input{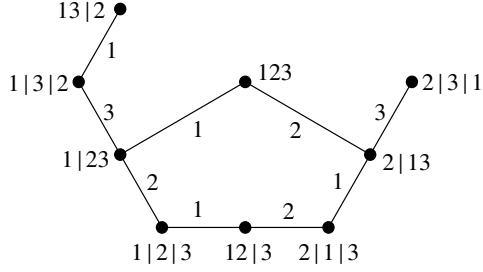}  

\caption{The graph $\Gamma_3(\Omega,V)$, for $V=\{2,3\}$,
  $\Omega=\{2,2\s 3\}$.}
\label{fig:fig3}
\end{figure}

Note how this relates to our previously used notations:
$\gn(V)=\gn(\emptyset,V)$. In line with thinking about the set $V$ as
a~forbidden set, we think about $\Omega$ as a~set of {\it allowed}
prefixes. See Figure~\ref{fig:fig3} for an example.

The next proposition states a~simple, but important property of
$\pre_V(\sigma)$.

\begin{prop} \label{pr:st1}
Let $\sigma$ be some full $[n]$-tuple, and let $V$ be any subset of
$[n]$. Assume that we are given $x\in F(\sigma)$, such that $x\notin
V$. Then, flipping with respect to $x$ does not change the $V$-prefix,
in other words, we have
\begin{equation}\label{eq:st1}
\pre_V(\sigma)=\pre_V(\cf(\sigma,x)).
\end{equation}
\end{prop}

\pr Assume $\sigma=A_1\s\dots\s A_k\s A_{k+1}\s\dots\s A_t$, such that
$A_1\s\dots\s A_k=\pre_V(\sigma)$, so for $1\leq i\leq k$, we have
$A_i\subseteq V$, and furthermore $A_{k+1}\not\subseteq V$. Since
$x\not\in V$, we have $x\in A_{k+1}\cup\dots\cup A_t$, so we can pick
$k+1\leq l\leq t$, such that $x\in A_l$. We now consider different
cases.

First, if $l\geq k+2$, then 
\[\cf(\sigma,x)=A_1\s\dots\s A_k\s A_{k+1}\s B_{k+2}\s\dots\s B_{\tilde t},\] 
for some sets $B_{k+2},\dots,B_{\tilde t}$, where $\tilde t=t+1$ or
$\tilde t=t-1$. Clearly, we then have
$\pre_V(\cf(\sigma,x))=A_1\s\dots\s A_k$.

Now assume $l=k+1$ and $|A_{k+1}|=1$, i.e., $A_{k+1}=x$.  Since $x\in
F(\sigma)$, we have $k+2\leq t$. The Case~2 of
Definition~\ref{df:flip} applies, and we have
\[\cf(\sigma,x)=A_1\s\dots\s A_k\s x\cup A_{k+2}\s A_{k+3}\s\dots\s
A_t.\] Hence again $\pre_V(\cf(\sigma,x))=A_1\s\dots\s A_k$, since
$x\notin V$.

Finally, assume $l=k+1$, and $|A_{k+1}|\geq 2$. The Case~1 of
Definition~\ref{df:flip} applies, and we have
\[\cf(\sigma,x)=A_1\s\dots\s A_k\s x\s A_{k+1}\sm x\s
A_{k+2}\s\dots\s A_t.\] Since $x\notin V$, we get
$\pre_V(\cf(\sigma,x))=A_1\s\dots\s A_k$ here as well. \qed

\subsection{Standard matchings} \label{ssect:sm}

$\,$ 

\nin We shall now describe a set of partial matchings on the graphs
$\gn(\Omega,V)$, which we shall call the {\it standard matchings}. To
start with, note that formally a~matching is a~function $\mu$ defined
on some of the vertices of $G$, which has vertices of $G$ as values,
and which satisfies the following conditions:
\begin{itemize}
\item if $\mu(\sigma)$ is defined, then the vertices $\sigma$ and
  $\mu(\sigma)$ are connected by an~edge, called the {\it matching
  edge};
\item when $\mu(\sigma)$ is defined, then $\mu(\mu(\sigma))$ is also
  defined and is equal to~$\sigma$.
\end{itemize}

When we consider matchings in the specific case of the flip graphs, we
can record the labels of the matching edges. Assuming $\mu(\sigma)$ is
defined, we let $\id_\mu(\sigma)$ denote the label of the matching
edge $(\sigma,\mu(\sigma))$; it is uniquely determined by the identity
$\mu(\sigma)=\cf(\sigma,\id_\mu(\sigma))$.  By~\eqref{eq:flfl}, we
have $\id_\mu(\mu(\sigma))=\id_\mu(\sigma)$.

\begin{df}\label{df:height}
Let $V$ be any subset of $[n]$, and let $R=(x_1,\dots,x_d)$ be an
order on its complement $[n]\sm V$; in particular, $d=n-|V|$. Assume
$\sigma=A_1\s\dots\s A_t$ is a~full $[n]$-tuple. We set $h_R(\sigma)$
to be the~index $1\leq h\leq d$, such that $\sigma=A_1\s\dots\s A_k\s
x_{h+1}\s\dots\s x_d$, and $A_k\neq x_h$.  Clearly, if such an~index
exists, it is unique. If it does not exist, we have
$\sigma=A_1\s\dots\s A_k\s x_1\s\dots\s x_d$, in which case we set
$h_R(\sigma):=0$. We call $h_R(\sigma)$ the {\bf height} of $\sigma$
with respect to~$R$.
\end{df}

By Definition~\ref{df:height}, we have $0\leq h_R(\sigma)\leq d$,
where $d=n-|V|$. The maximum $d$ is achieved if and only if $A_t\neq
x_d$. The full $[n]$-tuples of height $0$ with respect to some fixed
order $R$ are called {\it critical with respect to $R$.}

\begin{rem} \label{rm:crit}
The critical full $[n]$-tuples all begin by some full $V$-tuple,
followed by the full $([n]\sm V)$-tuple $x_1\s\dots\s x_d$.
\end{rem}
 
We now have the necessary terminology to define the standard
matchings.

\begin{df}\label{df:mr}
Assume $V$ is a~subset of $[n]$, and $R$ is an order on its complement
$[n]\sm V$.  We define a~partial matching on the vertices of $\gn$,
denoted by $\mu_R$.  For an arbitrary full $[n]$-tuple $\sigma$, set
$h:=h_R(\sigma)$. If $h\neq 0$, we set
\begin{equation}\label{eq:mu}
\mu_R(\sigma):=\cf(\sigma,x_h),
\end{equation}
else $\mu_R(\sigma)$ is undefined.  We call $\mu_R$ the {\bf standard
  matching} associated to $R$.
\end{df}

Note, that in the above Definition~\ref{df:mr}, we might as well ask
$V$ to be a~proper subset of $[n]$. This is because the case $V=[n]$
is rather degenerate: any $R$ is an empty order, and so the standard
matching $\mu_R$ is empty as well, with all vertices being critical
with respect to~$R$. We refer the reader to~\cite[Figure 3,
  p.\ 173]{wsb6}, for the illustration of the standard matching for
$n=3$.

\begin{prop} \label{pr:st2}
Whenever $V$ is a~subset of $[n]$, and $R$ is any order on $[n]\sm V$,
the partial matching $\mu_R$ on the set of vertices of $\Gamma_n$ is
well-defined. Furthermore, when $\mu_R(\sigma)$ is defined, we have
\begin{equation} \label{eq:st2-1}
\pre_V(\sigma)=\pre_V(\mu_R(\sigma))
\end{equation}
and 
\begin{equation} \label{eq:st2-2}
h_R(\sigma)=h_R(\mu_R(\sigma)).
\end{equation}
\end{prop}
\pr To say that $\mu_R$ is well-defined is equivalent to the following
statements:
\begin{enumerate}
\item[(1)] if $\mu_R(\sigma)$ is defined, then $\sigma$ and
  $\mu_R(\sigma)$ are connected by an edge;
\item[(2)] $\mu_R(\mu_R(\sigma))$ is also defined;
\item[(3)] $\mu_R(\mu_R(\sigma))=\sigma$.
\end{enumerate} 
The statement (1) is obvious, since $\mu_R(\sigma)$ is a~certain flip
of~$\sigma$. To verify the rest, assume $R=(x_1,\dots,x_d)$, and
$\sigma=A_1\s\dots\s A_k\s x_{h+1}\s\dots\s x_d$, where
$h=h_R(\sigma)$. Since $\mu_R(\sigma)$ is defined, we have
$h_R(\sigma)>0$, and $A_k\neq x_h$. By Definition~\ref{df:mr}, we have
$\mu_R(\sigma)=\cf(\sigma,x_h)$. By construction, we have $x_h\notin
V$, so by Proposition~\ref{pr:st1} we get
$\pre_V(\mu_R(\sigma))=\pre_V(\cf(\sigma,x_h))=\pre_V(\sigma)$,
and~\eqref{eq:st2-1} is proved.

Pick $1\leq l\leq k$, such that $x_h\in A_l$. Assume first $|A_l|=1$,
i.e., $A_l=x_h$. In this case we must have $l\leq k-1$. If $l\leq
k-2$, then
\[\mu_R(\sigma)=A_1\s\dots\s A_{l-1}\s x_h\cup A_{l+1}\s A_{l+2}\s\dots\s 
A_k\s x_{h+1}\s\dots\s x_d,\] and, since $x_h\neq A_k$, we get
$h_R(\sigma)=h_R(\mu(\sigma))$. If $l=k-1$ instead, we have
\[\mu_R(\sigma)=A_1\s\dots\s A_{k-2}\s x_h\cup A_k\s
 x_{h+1}\s\dots\s x_d,\] 
and, since $x_h\neq x_h\cup A_k$, we again
get $h_R(\sigma)=h_R(\mu(\sigma))$.

Assume now that $|A_l|\geq 2$. In this case $\mu_R(\sigma)$ is
obtained from $\sigma$ by splitting off the element~$x_h$. If $l\leq
k-2$, then
\[\mu_R(\sigma)=A_1\s\dots\s A_{l-1}\s x_h\s A_l\sm x_h\s A_{l+1}\s\dots\s 
A_k\s x_{h+1}\s\dots\s x_d,\] and, since $x_h\neq A_k$, we get
$h_R(\sigma)=h_R(\mu(\sigma))$.  Finally, if $l=k-1$, we have
\[\mu_R(\sigma)=A_1\s\dots\s A_{k-1}\s x_h\s A_k\sm x_h\s 
 x_{h+1}\s\dots\s x_d,\] 
and, since $x_h\neq A_k\sm x_h$, we again
get $h_R(\sigma)=h_R(\mu(\sigma))$.

The equality~\eqref{eq:st2-2} has now been proved for all $\sigma$.
In particular, if $h_R(\sigma)\neq 0$, then $h_R(\mu_R(\sigma))\neq 0$,
so $\mu_R(\mu_R(\sigma))$ is defined.

Finally, the equalities \eqref{eq:flfl}, \eqref{eq:st2-2}, and
definition of $\mu_R$, combine to
\[\mu_R(\mu_R(\sigma))=\cf(\cf(\sigma,x_h),x_h)=\sigma,\]
finishing the proof of the proposition.
\qed

\vskip5pt

Note that Proposition~\ref{pr:st2}, including the
equality~\eqref{eq:st2-1}, implies that $\mu_R$ restricts to a~partial
matching on $\gn(\Omega,V)$, for any~$\Omega$. The next theorem states
the key properties of standard matchings in flip graphs.

\begin{thm}\label{thm:main_match} $\,$
\begin{enumerate}
\item[(1)] Let $R=(x_1,\dots,x_n)$ be an~arbitrary order on the set
  $[n]$. The standard matching $\mu_R$ on $\gn$ is near-perfect, it
  has a~unique critical vertex, indexed by $x_1\s\dots\s x_n$.
\item[(2)] Let $V$ be an arbitrary non-empty subset of $[n]$, and let
  $R$ be any order on $[n]\sm V$. The associated standard matching
  $\mu_R$ on $\gn(V)$ is perfect.
\item[(3)] Assume $V$ is a non-empty subset of $[n]$,
  $R=(x_1,\dots,x_d)$ an order on $[n]\sm V$, and $\Omega$ a~family of
  $V$-tuples. The associated standard matching $\mu_R$ is a~partial
  matching on $\gn(\Omega,V)$, with critical vertices of the form
  $A_1\s\dots\s A_k\s x_1\s\dots\s x_d$, where $A_1\s\dots\s A_k$ is
  a~full $V$-tuple in $\Omega$. In particular, if $\Omega$ has no full
  $V$-tuples, then $\mu_R$ is a~perfect matching on $\gn(\Omega,V)$.
\end{enumerate} 
\end{thm}

\pr (1) We have $V=\emptyset$, hence $\pre_V(\sigma)=\emptyset$, for
all $\sigma$. By Remark~\ref{rm:crit} all critical vertices are
indexed by concatenations of full $V$-tuples with $x_1\s\dots\s x_d$,
where $d=n-|V|$.  Here that description reduces to the existence of
a~single critical vertex $x_1\s\dots\s x_n$.

(2) Let $\sigma$ be a~vertex of $\Gamma_n$ which is critical with
respect to $\mu_R$. We have $\sigma=A_1\s\dots\s A_k\s x_1\s\dots\s
x_d$, where $A_1\s\dots\s A_k$ is a~full $V$-tuple. Since $V\neq
\emptyset$, we have $k\geq 1$ and $A_1\subseteq V$. By
Definition~\ref{df:gnv} this vertex does not belong to $\gn(V)$, so
$\gn(V)$ has no critical vertices. Clearly, this is the same as to say
that $\mu_R$ restricts to a~perfect matching on the vertices
of~$\gn(V)$. Identical argument shows the more general statement (3)
as well.  
\qed


\section{Conductivity in the flip graphs} \label{sect:5}

\subsection{Previous work} $\,$

\noindent
While the standard matchings defined in subsection~\ref{ssect:sm} are
very useful, they do not always yield perfect matchings in the
situations we will be interested in. It is therefore practical to have
a procedure to modify a~partial matching so as to decrease the number
of critical vertices. To start with, let us recall the following
additional terminology from graph theory.

\begin{df}
Assume we are given a matching on a~graph $G$. An~edge path is called
{\bf alternating} if its edges are alternating between matching and
non-matching ones. It is called {\bf properly alternating} if, in
addition, it starts and ends either with a~matching edge, or with
a~critical vertex.

A~properly alternating path is called {\bf augmenting} if it starts
and ends with critical vertices, it is called {\bf non-augmenting} if
it starts and ends with matching edges, and, finally, it is called
{\bf semi-augmenting} if it is neither augmenting nor non-augmenting.
\end{df}

Next definition describes a~classical technique for modifying
matchings.

\begin{df}\label{df:mdef}
Assume we are given a matching $\mu$ on a~graph $G$, and a~properly
alternating non-self-intersecting edge path $\gamma$. We define
$D(\mu,\gamma)$ as a~new matching on $G$ consisting of all edges from
$\mu$ which do not belong to $\gamma$ together with all edges from
$\gamma$ which do not belong to~$\mu$.
\end{df}

When trying to modify a matching one is looking for existence of such
properly alternating non-self-intersecting edge paths~$\gamma$. It
turns out that when the underlying graph is bipartite, the condition
for the path to be non-self-intersecting can be dropped.

\begin{rem}\label{rem:interpath}
Assume $G$ is a~bipartite graph, $\mu$ is a~matching on $G$, $v$ and
$w$ are different vertices of $G$, and $\gamma$ is a~properly
alternating edge path from $v$ to $w$. Then there exists a~properly
alternating non-self-intersecting edge path from $v$ to~$w$.
\end{rem}
\pr If $\gamma$ does have self-intersections, then it contains
cycles. Any such cycle is of even length, since the graph is
bipartite. Deleting a cycle of even length from a~properly alternating
edge path yields another properly alternating edge path. If we keep
removing the cycles, we will eventually make our properly alternating
edge path non-self-intersecting.\qed \vskip5pt

Assume $\gamma$ is a~semi-augmenting path with endpoints $v$ and $w$,
where $v$ is a~critical vertex, and $w$ is not. It is easy to see that
the set of critical vertices with respect to $D(\mu,\gamma)$ is
obtained by taking the critical vertices with respect to $\mu$, and
then replacing $v$ with~$w$. For this reason, we shall intuitively
view the process of replacing $\mu$ with $D(\mu,\gamma)$ as {\it
  transporting $v$ to $w$ along the path~$\gamma$}. We think of the
corresponding property of the graph as its {\it conductivity}.  The
following result has been proved in \cite{wsb6}.

\begin{prop} \label{prop:cond} 
(\cite[Theorem 5.11, Theorem 5.12]{wsb6}) $\,$

\noindent
Assume $n$ is an~arbitrary natural number.
\begin{itemize}
\item The graph $\gn$ is a~bipartite graph with a~unique bipartite
  decomposition~$(A,B)$ such that $|A|=|B|+1$. For any vertex $v\in A$
  there exists a perfect matching on $\gn\sm v$.
\item Assume $[n]\supset V\neq\emptyset$, then the graph $\gn(V)$ is
  a~bipartite graph with a~bipartite decomposition~$(A,B)$ such that
  $|A|=|B|$. If furthermore $|V|\leq n-2$, then for any vertices
  $v\in A$, $w\in B$, there exists a perfect matching on
  $\gn\sm\{v,w\}$.
\end{itemize}
\end{prop}

The proof of the first part of Proposition~\ref{prop:cond} in
\cite{wsb6} was based on the fact that given a~near-perfect matching
(recall Definition~\ref{df:perfm}) of $\gn$ with a~critical vertex
$v\in A$, for any other vertex $w\in A$ there would exist
a~semi-augmenting path from $v$ to $w$. For the proof of the second
part, we constructed in \cite{wsb6} non-augmenting paths for any pair
of vertices $v\in A$ and $w\in B$.

Rather than directly generalizing the techniques used in~\cite{wsb6}
to prove Proposition~\ref{prop:cond}, we take a~slightly different
approach. Namely, instead of seeking to connect any pair of arbitrary
vertices, we single out a~special group of vertices, which we call
{\it connectors} and only try to conduct between them.

\begin{df}
Any vertex of $\gn$, which is indexed by a~full $[n]$-tuple
$a_1\s a_2\s\dots\s a_n$ is called an {\bf $a_n$-connector of the
  first type}, whereas any vertex indexed by a~full $[n]$-tuple
$\{a_1,a_2\}\s a_3\s\dots\s a_n$ is called an {\bf $a_n$-connector of
  the second type}.

Given a~connector $\tau=A_1\s\dots\s A_t$ of any of the two types,
and a~full $[n]$-tuple $\sigma$, we say that $\tau$ is {\bf proper}
with respect to $\sigma$, if $A_1\notin V(\sigma)$, in other words,
$\tau$ is a~well-defined vertex of $\gn(V(\sigma))$. 
\end{df}

In the rest of this section we assume that $n\geq 5$, and that
$n-1\geq |V|\geq 1$.

\subsection{Conductivity in $\gn(V,\Omega)$ when $\Omega$ has no 
full $V$-tuples} $\,$
\label{ssect:5}

\noindent Assume we are given a~family of $V$-tuples $\Omega$ which
has no full $V$-tuples. In this case, according to
Theorem~\ref{thm:main_match}(3) the standard matching associated to
any order is perfect.

%

\begin{lm}\label{lm:c1}
Let $\sigma\in\gn(V,\Omega)$, $\sigma=a_1\s\dots\s a_n$, be
an~arbitrary $a_n$-connector of the first type, such that $a_1\notin
V$.
\begin{enumerate}
\item[(1)] For any given element $f\neq a_1$, there exists an order
  $R$ on $[n]\sm V$, and an~edge path $p$ in $\gn(V,\Omega)$, which is
  non-augmenting with respect to $\mu_R$, starting from $\sigma$, and
  terminating at some $f$-connector of the second type $\tau=\{a_1,
  y_2\}\s\dots\s y_{n-1}\s f$.
\item[(2)] If $|V|\leq n-2$, there exists an order $R$ on $[n]\sm V$,
  and an~edge path $p$ in $\gn(V,\Omega)$, which is non-augmenting
  with respect to $\mu_R$, starting from $\sigma$, and terminating at
  some $a_1$-connector of the second type $\tau=\{y_1,y_2\}\s\dots\s
  y_{n-1}\s a_1$, with $\{y_1,y_2\}\not\subseteq V$.
\end{enumerate}
\end{lm}

\pr After renaming, we can assume without loss of generality that
$V=\{1,\dots,d\}$, where $d\leq n-1$, and that $a_1=n$, i.e.,
$\sigma=n\s a_2\s\dots\s a_n$. We now set $R:=(d+1,\dots,n)$.

We start by proving~(1), i.e., we are given $f\neq a_1$. Assume first
that $f=a_l$, for some $l\geq 3$. Using the alternating path
$\swap^I_k$ shown on left-hand side of the Figure~\ref{table:swapk} we
can swap the $k$-th and the $(k+1)$-st parts of our full $[n]$-tuple,
for any $k\geq 3$. We concatenate the paths $\swap^I_l$,
$\swap^I_{l+1}$, $\dots$, $\swap^I_{n-1}$ to obtain a~new alternating
path. This path ends at a~vertex of the form $\tilde\tau=n\s a_2\s
b_3\s\dots\s b_{n-1}\s f$, for some $b_3,\dots,b_{n-1}$, which is
an~$f$-connector of the first type. Set $\tau=\{n,a_2\}\s b_3\s\dots\s
b_{n-1}\s f$, and add the matching edge $(\tilde\tau,\tau)$ to our
path. We now have a~non-augmenting path between $\sigma$ and $\tau$,
with the latter being an~$f$-connector of the second type of the
required form. Thus the statement (1) is proved in this case.

Assume now that $f=a_2$, i.e., $\sigma=n\s f\s a_3\s\dots\s a_n$. In
this case we first follow the somewhat more complicated alternating
path $\swap^I_2$ shown on the left-hand side of the
Figure~\ref{table:swap2}, and then proceed as in the case $l\geq 3$,
by concatenating the alternating paths $\swap_3^I$, $\dots$,
$\swap_{n-1}^I$. Again, we will end up with a~non-augmenting path
between $\sigma$ and the $f$-connector of the second type
$\tau=\{n,a_3\}\s a_4\s\dots\s a_n\s a_2$. Note, that we use here the
fact that $n\geq 5$, implying $n-1>3$.  This finishes the proof
of~(1).

To prove (2) assume now that $|V|\leq n-2$, in particular, we have
$n-1\notin V$. Let $l\geq 2$ be such that $a_l=n-1$. If $l\geq 3$, we
start by concatenating the alternating paths $\swap^I_{l-1}$,
$\swap^I_{l-2}$, $\dots$, $\swap^I_2$, to arrive at the vertex of the
form $n\s n-1\s b_3\s\dots\s b_n$, for some $b_3,\dots,b_n$; if $l=2$
then we are at that vertex to start with. Note, that the alternating
paths $\up^I_k$, for $1\leq k\leq n-1$, allow in certain situations to
move the element $n$ from being the $k$-th set of our full $[n]$-tuple
to being its $(k+1)$-st set. We now concatenate the alternating paths
$\up^I_1$, $\up^I_2$, $\dots$, $\up^I_{n-1}$ to arrive at the vertex
$\tilde\tau= n-1\s b_3\s\dots\s b_n\s n$. To finish, set
$\tau=\{n-1,b_3\}\s\dots\s b_n\s n$, and add the matching edge
$(\tilde\tau,\tau)$ to our path. We now have a~non-augmenting path
between $\sigma$ and $\tau$, where $\tau$ is a~$a_1$-connector of the
second type satisfying the desired conditions. This finishes the proof
of Lemma~\ref{lm:c1}.  \qed

\begin{lm}\label{lm:c2}
Assume we are given a~connector of second type
$\sigma=\{a_1,a_2\}\s a_3\s\dots\s a_n$, such that $\{a_1,a_2\}\not\subseteq
V$, say $a_1\notin V$.
\begin{enumerate}
\item[(1)] For any $f\neq a_1$, there exists an order $R$ on $[n]\sm
  V$, and an~edge path $p$ in $\gn(V,\Omega)$, which is non-augmenting
  with respect to $\mu_R$, starting from $\sigma$, and terminating at
  some $f$-connector of the first type $\tau=a_1\s y_2\s\dots\s
  y_{n-1}\s f$.
\item[(2)] If $|V|\leq n-2$, there exists an order $R$ on $[n]\sm V$,
  and an~edge path $p$ in $\gn(V,\Omega)$, which is non-augmenting
  with respect to $\mu_R$, starting from $\sigma$, and terminating at
  some $a_1$-connector of the first type $\tau=y_1\s y_2\s\dots\s
  y_{n-1}\s a_1$, with $y_1\notin V$.
\end{enumerate}
\end{lm}

\pr Again, we can assume without loss of generality, that
$V=\{1,\dots,d\}$, where $d\leq n-1$, and that $a_1=n$, i.e.,
$\sigma=\{n,a_2\}\s a_3\s\dots\s a_n$. We now set $R:=(d+1,\dots,n)$.

First, we prove the statement (1). Assume $f=a_k$, $k\geq 3$. We can
concatenate the paths $\swap^{II}_k$, $\dots$, $\swap^{II}_{n-1}$,
which are shown in the right hand side of Figures~\ref{table:swapk},
and~\ref{table:swap2}. This will get us to the vertex
$\tilde\tau=\{n,y_2\}\s\dots\s y_{n-1}\s f$, for some
$y_2,\dots,y_{n-1}$. We set $\tau:=n\s y_2\s\dots\s y_{n-1}\s f$ and
note that $(\tilde\tau,\tau)$ is a~matching edge. Adding that edge to
the path which we have up to now yields a~non-augmenting path
connecting $\sigma$ with~$\tau$.

Let us now show (2). We have assumed that $|V|\leq n-2$, i.e.,
$n-1\notin V$. Here we have $\sigma=\{n,a_2\}\s a_3\s\dots\s a_n$, and
we pick index $k$ such that $a_k=n-1$. Assume first that $k\geq 3$. If
$k=3$, then we have $\sigma=\{n,a_2\}\s n-1\s a_4\s\dots\s a_n$. If
$k\geq 4$, then we can concatenate paths $\swap^{II}_{k-1}$,
$\swap^{II}_{k-2}$, $\dots$, $\swap^{II}_3$. This will yield
an~alternating path starting at $\sigma$ and terminating at
$\{n,a_2\}\s n-1\s a_4\s\dots\s a_n$. After that we concatenate with
the path $\specup^{II}$ shown on the left hand side of the
Figure~\ref{table:ups}. The obtained path terminates at the vertex
$\{n-1,a_2\}\s n\s a_4\s\dots$. Further, we concatenate with the
alternating paths $\up^{II}_3$, $\up^{II}_4$, $\dots$,
$\up^{II}_{n-1}$, see the Figure~\ref{table:up2k}, to arrive at the
vertex $\{n-1,a_2\}\s a_4\s\dots\s a_n\s n$. We finish by
concatenating with the matching edge between $\{n-1,a_2\}\s
a_4\s\dots\s a_n\s n$ and $\tau=n-1\s a_2\s a_4\s\dots\s a_n\s n$, to
obtain a~non-augmenting path from $\sigma$ to the appropriate
$n$-connector of the first type~$\tau$.

It remains to consider the case $k=2$, that is $\sigma=\{n-1,n\}\s
a_3\s a_4\s\dots$. In this situation we start with the alternating
path $\up^{II}_2$, see the right hand side of the
Figure~\ref{table:ups}, and arrive at the vertex $\{n-1,a_3\}\s n\s
a_4\s\dots$. We can then proceed just as in the case before with the
alternating paths $\up^{II}_3$, $\up^{II}_4$, $\dots$,
$\up^{II}_{n-1}$, followed up with the matching edge between
$\{n-1,a_3\}\s a_4\s\dots\s a_n\s n$ and $\tau=n-1\s a_3\s
a_4\s\dots\s a_n\s n$, to again obtain a~non-augmenting path from
$\sigma$ to the appropriate $n$-connector of the first
type~$\tau$. This is the last case to be considered and we have now
shown the statement~(2).  \qed

\begin{figure}[hbt]

  \input{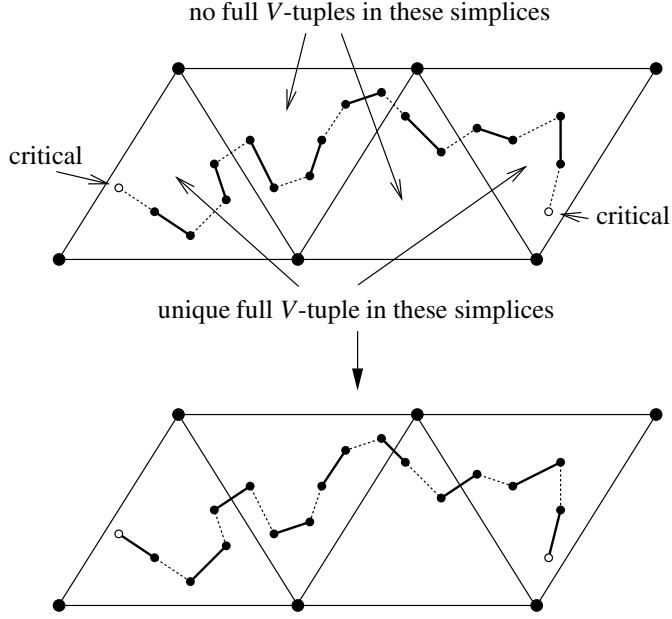}  

\caption{Concatenating augmenting paths.}
\label{fig:pext}
\end{figure}

Clearly, the Lemmata~\ref{lm:c1} and~\ref{lm:c2} allow us to extend
augmenting paths across $\gn^2$ as shown on the Figure~\ref{fig:pext}.

Note that if $|V|\leq n-2$, then there are $x$-connectors of both
types for all~$x$.  If $|V|=n-1$, then there are $x$-connectors of
both types if and only if $x\neq[n]\sm V$.  If $|V|=n-1$ and
$[n]=V\cup\{x\}$, then there are no $x$-connectors, but we also do not
need any.

\subsection{Conductivity in $\gn(V,\Omega)$ in some special cases} $\,$

\noindent Let us first consider the case when $\Omega$ has a~unique
full $V$-tuple. Again, according to Theorem~\ref{thm:main_match}(3)
the standard matching associated to any order has a~unique critical
vertex.

\begin{lm}
\label{lm:sc+}
Assume that the family $\Omega$ contains a~unique full $V$-tuple
$v_1\s\dots\s v_d$ together with all of its truncations. Assume
furthermore, that we are given some connector of the first type
$\tau=y_1\s\dots\s y_n$, such that $y_1\notin V$. Then there exists an
order $R$ on $[n]\sm V$, and an~edge path in $\gn(V,\Omega)$, which is
semi-augmenting with respect to $\mu_R$, and which connects the
critical vertex $\sigma$ to~$\tau$.
\end{lm}
\pr After suitable renaming we can assume, without loss of generality,
that $V=\{1,\dots,d\}$, and that the unique full $V$-tuple is
$1\s\dots\s d$. Furthermore, we can make sure that $y_1=n$ after that
renaming.  We now choose the order $R:=(d+1,\dots,n)$, hence the
unique, critical with respect to $\mu_R$, vertex is $\sigma=1\s
2\s\dots\s n$. We need to find a~semi-augmenting path from $\tau=n\s
y_2\s\dots\s y_n$ to~$\sigma$.

To start with, we can concatenate paths $\swap^I_k$, for $2\leq k\leq
n-1$ in an appropriate order, so as to obtain an~alternating path
starting at $\tau$ and terminating at $n\s 1\s 2\s\dots\s n-1$. After
this, we concatenate the paths $\up_1^I,\dots,\up_{n-1}^I$. Note, that
these paths lie within the graph $\gn(V,\Omega)$, since we assumed
that $\Omega$ contains all truncations of $1\s\dots\s d$. The total
path terminates at $\sigma$, which is exactly what we are looking
for. \qed

Let us now consider the~second special case. This time we assume that
$\Omega$ has three full $V$-prefixes: $(v_1\s v_2\s v_3\s\dots\s
v_d)$, $(\{v_1,v_2\}\s v_3\s\dots\s v_d)$, and $(v_1\s \{v_2,v_3\}
\s\dots\s v_d)$. In this case the standard matching $\mu_R$ associated
to any order has three critical vertices. We shall extend $\mu_R$ by
matching two of the critical vertices to each other. After this we
find an augmenting path from the third critical vertex to
a~$y$-connector, as in Lemma~\ref{lm:sc+}.

\begin{lm}
\label{lm:sc-}
Assume we are given set $V$ and a~family of $V$-tuples $\Omega$, which
contains three full $V$-tuples as above, together with all of their
truncations. Assume, furthermore, we are given $f\in [n]$, such that
$[n]\sm V\neq f$. Then there exists an order $R$ on $[n]\sm V$, such
that the standard matching $\mu_R$ can be extended by matching two of
the critical vertices to each other, and, furthermore, there exists
an~edge path in $\gn(V,\Omega)$, which is semi-augmenting with respect
to that extended matching, and which connects the remaining critical
vertex $\sigma$ to some $f$-connector of the second type
$\{y_1,y_2\}\s y_3\s\dots\s y_{n-1}\s f$, such that
$\{y_1,y_2\}\not\subseteq V$.
\end{lm}
\pr Again, after suitable renaming, we can assume, without loss of
generality, that $V=\{1,\dots,d\}$, and that the full $V$-tuples are
$(1\s 2\s 3\s\dots\s d)$, $(\{1,2\}\s 3\s\dots\s d)$, and
$(1\s\{2,3\}\s\dots\s d)$. Since $[n]\sm V\neq f$, we can pick
an~element of $[n]\sm V$ different from~$f$. Without loss of
generality we can make sure, that after renaming that element is
called $n$. We set $R:=(d+1,\dots,n)$, so the three critical vertices
are now $\sigma=\{1,2\}\s 3\s 4\s\dots\s n$, $\alpha_1=1\s 2\s 3\s
4\s\dots\s n$, and $\alpha_2=1\s\{2,3\}\s 4\s\dots\s n$. We extend the
standard matching $\mu_R$ by matching $\alpha_1$ with $\alpha_2$.

By our construction, $f\neq n$, and we set $\tau:=\{n,1\}\s 2\s\dots\s
f-1\s f+1\s\dots\s n-1\s f$. This is an~$f$-connector of the second
type satisfying necessary conditions, since $n\notin V$. We now
describe how to find a~semi-augmenting path from $\tau$
to~$\sigma$. To start with, we concatenate the paths
$\swap_{n-1}^{II}$, $\dots$, $\swap_{f+1}^{II}$, to arrive at the
vertex $\{n,1\}\s 2\s 3\s\dots\s n-1$. After this, we concatenate with
the path on Figure~\ref{table:star} to get to the desired
semi-augmenting path to~$\sigma$.  \qed

\section{Nodes}\label{sect:6}

\subsection{Definition of $n$-nodes of the $d$-th level} $\,$

\nin It is now time to define the {\it nodes}, which, after the flip
graphs, constitute the second main combinatorial concept of this
paper. On the geometric side the nodes correspond to vertices of
iterated chromatic subdivisions, while on distributed computing side
they correspond to local views of the processes.

\begin{df}\label{df:node}
Let $n$ and $d$ be arbitrary natural numbers. A~$d$-tuple
$v=\nu_1\ds\dots\ds\nu_d$ of coherent pairs of $[n]$-tuples
is called an {\bf $n$-node of the $d$-th level} if it
satisfies the following properties:
\begin{enumerate}
\item[(1)] $\col(\nu_i)=\carr(\nu_{i+1})$, for all $1\leq i\leq d-1$;
\item[(2)] $|\col(\nu_d)|=1$, in other words, there exists $S\subseteq
  [n]$ and $x\in S$, such that $\nu_d=(S,x)$.
\end{enumerate}
We set $\carr(v):=\carr(\nu_1)$, and call it the {\bf carrier} of~$v$;
we set $\col(v):=\col(\nu_d)$, and call it the {\bf color} of~$v$.

Finally, let $\cn_n^d$ denote the set of all $n$-nodes of the $d$-th level.
\end{df}

\begin{figure}[hbt]

  \input{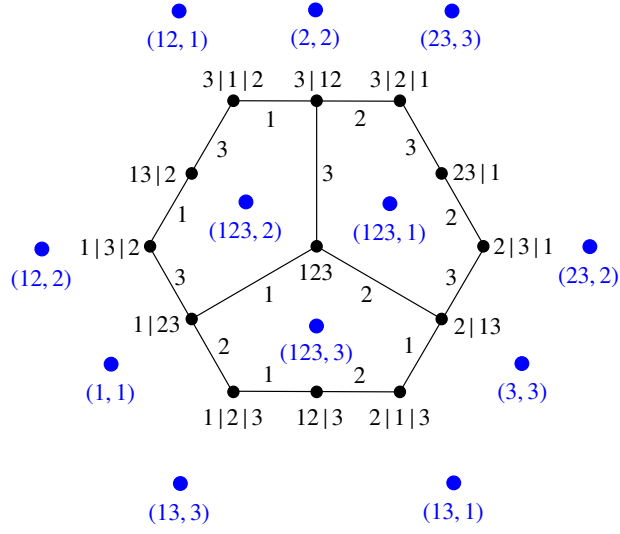}  

\caption{The $3$-nodes of the first level juxtaposed on $\Gamma_3$.}
\label{fig:fig4}
\end{figure}

The special cases $d=1$ and $d=2$ are the ones most used in this paper,
therefore it makes sense to unwind the Definition~\ref{df:node} to see
explicitly what it says for these values of~$d$.

\begin{itemize}
\item An {\it $n$-node of the first level} is simply a~pair $(S,x)$,
  where $S\subseteq[n]$ and $x\in S$.

\item An {\it $n$-node of the second level} is a pair $\sigma\ds\tau$,
  where $\sigma$ is a~coherent pair of $[n]$-tuples, and $\tau=(S,x)$ is
  an $n$-node of the first level, such that $\col(\sigma)=S$.
\end{itemize}

Assume we have a~bijective set map $\varphi:S\ra T$. Then we have an
induced map taking $S$-tuples to $T$-tuples, it is simply given by
$\varphi(A_1\s\dots\s A_t)=\varphi(A_1) \s\dots\s\varphi(A_t)$, that
is we apply $\varphi$ to each set separately.  In the same way, the
function $\varphi$ extends to coherent pairs of $S$-tuples, as well as
to tuples of coherent pairs of $S$-tuples.

In particular, assume $v$ is an $n$-node of $d$-th level, set
$S:=\supp(v)$, and assume we are given a~bijective map $\varphi:S\ra
T$.  Then $\varphi(v)$ is well-defined, it is an~$n$-node of $d$-th
level, and $\supp(\varphi(v))=T$.

\begin{df}
Let $v$ be an $n$-node of the $d$-th level, and let $\varphi$ be the
normalizer of $\carr(v)$. We call $\varphi(v)$ the {\bf normal form}
of~$v$.
\end{df}

\begin{df}\label{df:parent}
Let $n$ and $d$ be arbitrary natural numbers. Given an $n$-node of the
$(d+1)$-st level $v=v_1\ds\dots\ds v_{d+1}$, we define a~new $n$-node
$w=w_1\ds\dots\ds w_d$ of the $d$-th level as follows:
$w_d:=v_d\dar\col(v)$, $w_{d-1}:=v_{d-1}\dar\carr(w_d)$, $\dots$,
$w_1:=v_1\dar\carr(w_2)$.

We call the obtained node $w$ the {\bf parent} of $v$ and denote it by
$\parent(v)$.
\end{df}

Let us note a few special cases. If $d=1$, we have $v=v_1\ds (S,x)$,
and we set $\parent(v):=(T,x)=v_1\dar x$. If $d=2$, we have $v=v_1\ds
v_2\ds (S,x)$, and we set $\parent(v):=w_1\ds (T,x)$, where
$(T,x):=v_2\dar x$, and $w_1:=v_1\dar T$.

\begin{dcc}
The nodes are ``local views'' when $n$ processes run a~standard
protocol for $d$ rounds.
\end{dcc}

\subsection{Adjacency of nodes and vertices of the flip graphs} $\,$

\nin The $n$-nodes of the $d$-th level and vertices of $\gn^d$ are
related by means of adjacency. We start by giving the general
definition.

\begin{df}\label{df:adj}
Let $n$ and $d$ be arbitrary natural numbers. Assume we are given an
$n$-node $v=\nu_1\ds\dots\ds\nu_d$ of the $d$-th level and a~vertex
$\sigma=\sigma_1\ds\dots\ds\sigma_d$ of $\gn^d$.  We say that $v$ and 
$\sigma$ are {\bf adjacent} if $\nu_d=\sigma_d\dar\col(v)$ and
\begin{equation} \label{eq:adj}
\nu_i=\sigma_i\dar\carr(\nu_{i+1}),
\end{equation}
for all $i=1,\dots,d-1$.
\end{df}

It is again instructive to describe explicitly the cases $d=1$ and
$d=2$. When $d=1$, we have $v=(S,x)$, and $\sigma=A_1\s\dots\s
A_t$. Let $k$ be the index $1\leq k\leq t$, such that $x\in A_k$. Then
the vertex $\sigma$ and the node $(S,x)$ are {\it adjacent} if and
only if $S=A_1\cup\dots\cup A_k$.

On the other hand, when $d=2$, we have an $n$-node of the second level
$v=(\alpha,\beta)\ds\ab (S,x)$, and a~vertex of $\gn^2$,
$\sigma=\sigma_1\ds\sigma_2$.  Let $\sigma_2=B_1\s\dots\s B_q$, and
let $k$ be the index $1\leq k\leq q$, such that $x\in B_k$. We say
that the vertex $\sigma$ and the node $v$ are {\bf adjacent} if the
following conditions are satisfied:
\begin{itemize}
\item $S=B_1\cup\dots B_k$;
\item $(\alpha,\beta)=\sigma_1\dar S$.
\end{itemize}

It is easy to see that every vertex of $\gn^d$ is adjacent to exactly
$n$ nodes of $d$-th level. This is because, once a~vertex of $\gn^d$
is fixed, the color of the node $v$ defines the node $v$ uniquely by
means of equations~\eqref{eq:adj}.

\begin{dcc}
The adjacency encodes correspondence between local views and global
executions. Namely, an~$n$-node of the $d$-th level $v$ is adjacent to
a~vertex $\sigma$ of $\gn^d$ if and only if the local view of
a~process encoded by~$v$ is a~view contained in the execution of the
$d$-round protocol encoded by~$\sigma$.
\end{dcc}

\subsection{Node labelings} $\,$

\nin The main result of this paper is a~construction of a~function on
the set of the nodes satisfying certain constraints.

\begin{df}
Assume we are given arbitrary natural numbers $n$ and $d$. A~labeling
of the $n$-nodes of the $d$-th level, or simply a~{\bf node labeling},
is a~function $\lambda:\cn_n^d\ra\Lambda$, where $\Lambda$ is an
arbitrary set. A {\bf binary node labeling} is a function
$\lambda:\cn_n^d\ra\{0,1\}$.
\end{df}

Unless explicitly stated otherwise, all our node labelings will be
binary, so we will frequently omit that word.

\begin{df}\label{df:supp}
An $n$-node $v$ is called {\bf internal} if $\carr(v)=[n]$. Any node
which is not internal is called a~{\bf boundary node}.
\end{df}

Note that in particular $\supp(v)\supseteq\carr(\nu_i)$, for all
$i=1,\dots,d$, so the carrier of $v$ is sort of a~universe, containing
all the sets needed to define~$v$.

It is easy to rephrase Definition~\ref{df:supp} in the special cases
$d=1$ and $d=2$. An~$n$-node of the first level $(S,x)$ is internal if
and only if $S=[n]$, indeed its carrier is simply given by~$S$. On the
other hand, an~$n$-node of the second level $(A_1\s\dots\s A_t,
B_1\s\dots\s B_t) \ds (S,x)$ is internal if and only if
$A_1\cup\dots\cup A_t=[n]$.

\begin{df}
A binary node labeling $\lambda:\cn_n^d\ra\{0,1\}$ is called {\bf blank}
if $\lambda(v)=0$ whenever $v$ is an~internal node.
\end{df}

We want to look at the blank binary node labelings which satisfy a
certain condition on the boundary. 

\begin{df}\label{df:comp}
The binary node labeling $\lambda:\cn_n^d\ra\{0,1\}$ is called {\bf
  compliant} if the following property is satisfied. Assume we are
given two $n$-nodes of the $d$-th level, say $v$ and $w$, such that
$|\supp(v)|=|\supp(w)|$. Let $\varphi:\supp(v)\ra\supp(w)$ be the
unique order-preserving bijection, and assume furthermore that
$w=\varphi(v)$. Then we have $\lambda(v)=\lambda(w)$.
\end{df}

Note, that when $|\supp(v)|=n$, i.e., when $v$ is an internal node,
the condition in Definition~\ref{df:comp} is empty, since $\varphi$
must be the identity map. Thus being compliant is really a~condition
on the boundary nodes in~$\cn_n^d$.

\begin{df}
Assume we are given a~binary node labeling
$\lambda:\cn_n^d\ra\{0,1\}$.  A vertex $\sigma\in V(\gn^d)$ is called
{\bf $0$-monochromatic} if $\lambda(w)=0$ for any node $w\in\cn_n^d$
which is adjacent to~$\sigma$. Analogously, a~vertex $\sigma\in
V(\gn^d)$ is called {\bf $1$-monochromatic} if $\lambda(w)=1$ for any
node $w\in\cn_n^d$ which is adjacent to~$\sigma$.
\end{df}

The next definition describes the most important class of node labelings
in this paper.

\begin{df}
A binary node labeling is called {\bf symmetry breaking} if it is
compliant and does not have monochromatic vertices.
\end{df}

\begin{prop}
Any vertex of $\gn^d$ is adjacent to some internal node. In
particular, $\gn^d$ has no $1$-monochromatic vertices under a~blank
binary labeling.
\end{prop}

\pr Assume $\sigma=\sigma_1\ds\dots\ds\sigma_d$ is a~vertex of
$\gn^d$. Let us say $\sigma_d=A_1\s\dots\s A_t$. Take any $x\in A_t$,
and let $v$ be the unique $n$-node of $d$-th level whose id is $x$ and
which is adjacent to $\sigma$. It is easy to see, using
Definition~\ref{df:adj}, that $v=\sigma_1\ds\dots\ds\sigma_{d-1}\ds
([n],x)$.  Clearly, this node is internal. Finally, this implies that
we have no $1$-monochromatic vertices, since a~blank binary node
labeling evaluates to $0$ on any internal node.  \qed

\begin{df}
Given a binary node labeling $\lambda:\cn_n^d\ra\{0,1\}$, let $\cm_\lambda$
denote the subgraph of $\gn^d$ induced by the $0$-monochromatic vertices.

Furthermore, for any natural number $q<d$, and whenever $\sigma$ is
a~vertex of $\gn^q$, we let $\cm_\lambda(\sigma)$ denote the
intersection of $\gn^d(\sigma)$ with $\cm_\lambda$.
\end{df}

\begin{figure}[hbt]

  \input{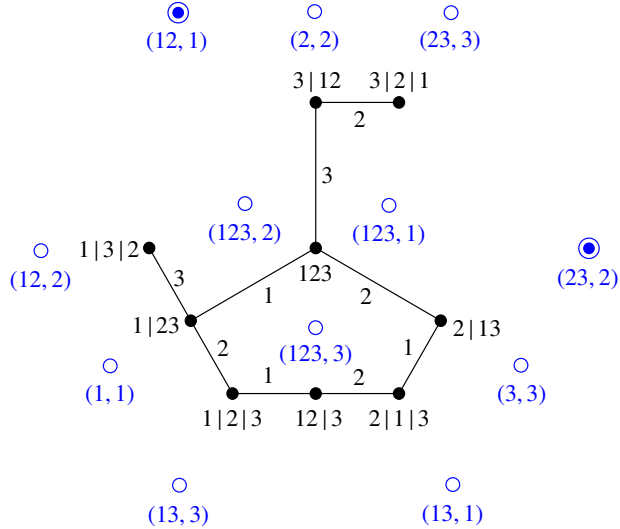}  

\caption{The graph $\cm_\lambda$ for the case when the only
  nodes labeled with $1$ are $(12,1)$ and $(23,2)$.}
\label{fig:fig5}
\end{figure}

When dealing with blank labelings we shall automatically have no
$1$-monochromatic vertices. Our next step will be to eliminate
$0$-monochromatic vertices as well, by looking for matchings on the
graph~$M_\lambda$.


\section{Sets of patterns}\label{sect:7}

\subsection{Definition and some specific sets of patterns}

\begin{df}
For an arbitrary natural number $n$, a~{\bf set of patterns} in $[n]$
is a~union $\cb_1\cup\dots\cup\cb_{n-1}$, where for each $1\leq k\leq
n-1$, $\cb_k$ is some set of the $k$-nodes of the first level.
\end{df}

As an example we consider the set of patterns which has been
instrumental in our previous work, \cite{wsb6}, when we analyzed the
case $n=6$. Rephrasing the construction from \cite{wsb6} in the
language of this paper, yields the following set of patterns in $[6]$:
\[\begin{array} {lcl}
\cb_1&=&\{(\{1\},1)\}, \\
\cb_2&=&\{(\{1\},1),(\{1,2\},2)\}, \\
\cb_3&=&\{(\{1\},1),(\{1,2\},1),(\{1,2\},2),(\{1,2,3\},2),(\{1,2,3\},3)\}.
\end{array}\]

The case-by-case analysis which we did in \cite{wsb6} can be derived
from the general structure results which we prove in this paper.

For future reference we define a certain special set of patterns.
Assume $n$ is an arbitrary natural number and
$\fatx=(x_1,\dots,x_{n-1})$ is a~vector, where $x_1,x_2\in\{0,1\}$,
$x_i\in\{-1,0,1\}$, for all $3\leq i\leq n-1$. The set of patterns
$\cb_\fatx=(\cb_1,\dots,\cb_{n-1})$ is now defined by the following
rule:
\[\cb_k:=
\begin{cases}
\cp^+_k&\text{ if }x_k=1;\\
\cp^-_k&\text{ if }x_k=-1;\\
0&\text{ otherwise;}
\end{cases}\]
for all $k=1,\dots,n-1$, where we set
\[\cp_k^+:=\{(\{1\},1),(\{1,2\},2),\dots,(\{1,\dots,k\},k)\},
\text{ for all } 1\leq k\leq n-1,\]
\[\cp_k^-:=\cp_k^+\cup\{(\{1,2\},1),(\{1,2,3\},2)\},
\text{ for all } 3\leq k\leq n-1.\] 
We are not aware of any nice interpretation or intuition behind the
sets of patterns $\cp^+_k$ and $\cp^-_k$. For us, these are technical 
constructions, which are needed to emulate the appearance of signs
in the solutions of the associated Diophantine equations.

We say that the set of patterns $\cb_\fatx$ {\it is associated} to the
vector~$\fatx$.

\begin{figure}[hbt]

  \input{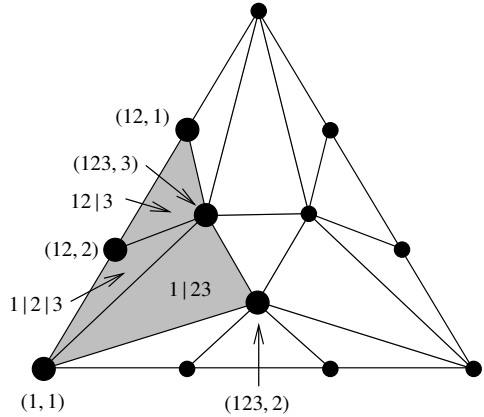}  

\caption{The set of patterns $\cp^-_3$ viewed geometrically.}
\label{fig:p-3}
\end{figure}

\subsection{Node labeling associated to sets of patterns}

\begin{df}\label{df:lb}
Whenever $\cb$ is some set of patterns in $[n]$, we define a~certain
binary node labeling $\lambda_\cb:\cn_n^2\ra\{0,1\}$, which we say is
{\bf associated} to~$\cb$. Pick $v\in\cn_n^2$, $v=(A_1\s\dots\s
A_t,B_1\s\dots\s B_t)\ds(S,x)$. One of the following 3 cases must
occur.

\nin {\bf Case 1.} The $n$-node $v$ is internal. In that case, we set
$\lambda_\cb(v):=0$.

\nin {\bf Case 2.} The $n$-node $v$ is a~boundary node, such that
$t\geq 2$. In that case, we set $\lambda_\cb(v):=1$.

\nin {\bf Case 3.} The $n$-node $v$ is a~boundary node, such that
$t=1$, in other words, we can write $v=(A,S)\ds(S,x)$, where $A\neq
     [n]$.  Let $\varphi$ be the normalizer of~$A$. We set
\[\lambda_\cb(v):=\begin{cases}
0,& \text{ if } (\varphi(S),\varphi(x))\in\cb; \\
1,& \text{ otherwise.}
\end{cases}\] 
\end{df}

Definition~\ref{df:lb} provides us with a~method of generating a~large
family of blank and compliant binary node labelings as the next
proposition shows.

\begin{prop}\label{prop:bcl}
The binary node labeling associated to an arbitrary set of patterns
is blank and compliant. 
\end{prop}

\pr Assume $\cb$ is some given set of patterns in $[n]$. The binary
node labeling $\lambda_\cb$ is set to be $0$ on the internal vertices
by definition, so it is blank. To see that it is also compliant, pick
two $n$-nodes of the second level, say $v$ and $w$, such that
$|\carr(v)|=|\carr(w)|$. Let $\varphi:\carr(v)\ra\carr(w)$ be the
unique order-preserving bijection, and assume that $w=\varphi(v)$; in
other words, if $v=(A_1\s\dots\s A_t,B_1\s\dots\s B_t)\ds(S,x)$, then
$w=(\varphi(A_1)\s\dots\s\varphi(A_t),\varphi(B_1)\s\dots\s\varphi(B_t))
\ds(\varphi(S),\varphi(x))$. 

If $v$ is an internal vertex, then $v=w$, so
$\lambda_\cb(v)=\lambda_\cb(w)=0$. Assume now $v$ is a~boundary
vertex.  If $t\geq 2$, then Case~2 of Definition~\ref{df:lb} applies
both to $v$ and to $w$, so we get $\lambda_\cb(v)=\lambda_\cb(w)=1$.
Assume finally $t=1$. Let $\psi$ be the normalizer of $A_1$, then
$\psi\circ\varphi^{-1}$ is a~normalizer of $\varphi(A_1)$. Note, that
$(\gamma(\varphi(S)),\gamma(\varphi(x)))=(\psi(S),\psi(x))$. In
particular $\gamma(\varphi(S),\varphi(x))\in\cb$ if and only if
$\psi(S,x)\in\cb$, implying that $\lambda_\cb(v)=\lambda_\cb(w)$ in
this final case as well.  \qed

\begin{df}\label{df:compos}
Let $\cb$ be an arbitrary set of $n$-nodes of the first level. We let
$\cp(\cb)$ denote the set of $[n]$-tuples $C_1\s\dots\s C_t$, such that
$(C_1\cup\dots\cup C_i,x)\in\cb$, for any $1\leq i\leq t$, and any
$x\in C_i$. We say that $\cp(\cb)$ consists of all $[n]$-tuples which 
can be {\bf composed} from~$\cb$.
\end{df}

Note, that if an~$[n]$-tuple $C_1\s\dots\s C_t$ can be composed then
then any of its truncations can be composed as well. 

\begin{simp}
The notion of being {\it composed} has an interesting simplicial
interpretation. Recall, that the $n$-nodes of first level correspond
to vertices of $\chi(\da^{n-1})$, which is the standard chromatic
subdivision of an~$(n-1)$-simplex. Given $\cb$ as in
Definition~\ref{df:compos}, we let $K_\cb$ denote the simplicial
complex induced by $\cb$, that is consisting of all simplices from
$\chi(\da^{n-1})$ whose vertices are in $\cb$. Call a~simplex of
$\chi(\da^{n-1})$ essential if it is contained in a~simplex of
$\da^{n-1}$ of the same dimension. Then $\cp(\cb)$ consists of
(indexes of) all essential simplices of $K_\cb$.
\end{simp}

As an example, we have
\[\cp(\cp_3^-)=\{1,1\s 2,12,1\s 2\s 3,1\s 23,12\s 3\}.\]
Simplicially, these correspond to the $6$ essential simplices on
Figure~\ref{fig:p-3}: $1$ vertex, $2$ edges, and $3$ triangles.

\begin{df}\label{df:pbk}
Assume we are given $\cb=\cb_1\cup\dots\cup\cb_{n-1}$ - a~set of
patterns in $[n]$, and $\sigma=A_1\s\dots\s A_t$ - a~full
$[n]$-tuple. We shall now give an algorithm producing a set of
$V$-tuples, where $V=[n]\sm A_t$. This set of $V$-tuples will be
denoted $\Omega(\cb,\sigma)$.

Pick some $1\leq l\leq t-1$. Let $\varphi:A_1\cup\dots\cup A_l\ra[k]$
be the normalizer of $A_1\cup\dots\cup A_l$. We define a~set of
$V$-tuples $\Omega_k$ by saying that a~$V$-tuple $C_1\s\dots\s C_q$
belongs to $\Omega_k$ if and only if
$\varphi(C_1)\s\dots\s\varphi(C_q)$ belongs to $\cp(\cb_k)$ and
$C_1\cup\dots\cup C_q\subseteq A_l$. We now set
$\Omega(\cb,\sigma):=\Omega_1\cup\dots\cup\Omega_{t-1}$.
\end{df}
 
\begin{rem}\label{rm:fp}
Let us note for future reference two important properties of
$\Omega(\cb,\sigma)$.
\begin{enumerate}
\item[(1)] The set  $\Omega(\cb,\sigma)$ is closed under taking
truncations.
\item[(2)] If $t\geq 3$, then $\Omega(\cb,\sigma)$ does not contain
any full $V$-tuples.
\end{enumerate}
\end{rem}
\pr Assume $C_1\s\dots\s C_q\in \Omega(\cb,\sigma)$. Picking $l$ as in
Definition~\ref{df:pbk} we see that $C_1\cup\dots\cup C_{q-1}\subseteq
A_l$, and $\varphi(C_1)\s\dots\s\varphi(C_{l-1})\in\cp(\cb_k)$; which
shows (1).  Property (2) follows from the condition that whenever
$C_1\s\dots\s C_q$ belongs to $\Omega_k$, we have $C_1\cup\dots\cup
C_q\subseteq A_l$, for some $1\leq l\leq t-1$.  \qed

\subsection{Flip graphs associated to sets of patterns} $\,$

\nin
The next theorem allows us to understand the combinatorial structure
of the subgraphs $\cm_{\lambda(\cb)}(\sigma)$.

\begin{thm}\label{thm:pat}
Assume $\cb=\cb_1\cup\dots\cup\cb_{n-1}$ is an arbitrary set of
patterns in $[n]$ and $\sigma=S_1\s\dots\s S_t$ is a~full
$[n]$-tuple.  We have an isomorphism
\begin{equation}\label{eq:pat}
\cm_{\lambda_\cb}(\sigma)\cong\gn(\Omega(\cb,\sigma),V(\sigma)),
\end{equation}
given by $(\sigma\ds\tau)\mapsto\tau$.
\end{thm}
\pr Let us take a~vertex $\sigma\ds\tau$ of the graph
$\cm_{\lambda_\cb}(\sigma)$ and show that $\tau$ is a vertex of
$\gn(\Omega(\cb,\sigma),V(\sigma))$.  Assume $\tau=T_1\s\dots\s T_q$,
and choose $1\leq k\leq q$ to be the minimal index such that
$T_{k+1}\cap S_t\neq\emptyset$. We need to show that $T_1\s\dots\s
T_k\in\Omega(\cb,\sigma)$.

Pick an arbitrary $1\leq i\leq k$, and $x\in T_i$. Set
$T:=T_1\cup\dots\cup T_i$. The node of the second level $w=(\sigma\dar
T)\ds(T,x)$ is obviously adjacent to the vertex $\sigma\ds\tau$, so
$\lambda_\cb(w)=0$. It cannot be internal since $T\cap S_t=\emptyset$,
so we assume it is a~boundary node. Since $\lambda_\cb(w)=0$, and $w$
is a~boundary node, we must have $w=(S,T)\ds(T,x)$, where $S\supseteq
T\ni x$. This means that there exists an index $1\leq d\leq t$, such
that $S_d\supseteq T$. Since this is true for any $i$, we get
$S_d\supseteq T_1\cup\dots\cup T_k$.  Furthermore, we have
$S=S_1\cup\dots\cup S_d$, and we set $m:=|S|$. Let $\varphi:S\ra[m]$
be the normalizer of $S$. Since $\lambda_\cb(w)=0$, we must have
$\varphi(T,x)\in\cb_m$, for all $x$, which means precisely that
$\varphi(T_1)\s\dots\s\varphi(T_k)\in \cp(\cb_k)$. By
Definition~\ref{df:pbk} we conclude that $T_1\s\dots\s T_k
\in\Omega(\cb,\sigma)$.

This argument can easily be reversed to show that for any vertex
$\tau$ in $\gn(\Omega(\cb,\sigma),V(\sigma))$, the vertex
$\sigma\ds\tau$ belongs to $\cm_{\lambda_\cb}(\sigma)$. Indeed, pick
$\tau=T_1\s\dots\s T_q\in\gn(\Omega(\cb,\sigma),V(\sigma))$, and let
$k$ be the minimal index such that $T_{k+1}\cap S_t\neq\emptyset$.  We
need to show that all the nodes adjacent to $(\sigma\ds\tau)$ have
a~label~$0$. This is clearly the cases for internal nodes, so let us
consider a~boundary node. This means we need to pick some $1\leq i\leq
k$, and some $x\in T_i$. The corresponding node is $w=(\sigma\dar
T)\ds(T,x)$, where $T=T_1\cup\dots\cup T_i$. Since $T_1\s\dots\s T_k
\in\Omega(\cb,\sigma)$, there exists $d$ such that $T\subseteq S_d$.
Setting $S:=S_1\cup\dots\cup S_d$, we get $w=(S,T)\ds(T,x)$. If
$\varphi$ is a~normalizer of~$S$, then
$(\varphi(T),\varphi(x))\in\cb$, and we conclude that the node $w$ has
label~$0$.

Finally, we get a~graph isomorphism, since graphs on both sides
of~\eqref{eq:pat} are induced by their respective sets of vertices.
\qed


\section{Sets of disjoint paths in $\gn$}\label{sect:8}

\begin{df} \label{df:nest}
A {\bf well-ordered pair} of sets is a~pair of sets $(S,T)$, such that
$\emptyset\neq S\subset T\subset [n]$, together with some fixed order
on the set~$T$, under which all the elements of~$S$ come before all
the other elements of~$T$. Two well-ordered pairs of sets $(S,T)$ and
$(S',T')$ are called {\bf nested} if either $S\subset S'\subset
T'\subset T$ or $S'\subset S\subset T\subset T'$.
\end{df}


For an arbitrary $S\subseteq[n]$, we let $\fatb_S$ denote the vertex
of $\gn$ indexed by $S\s [n]\sm S$.

\begin{df} 
Given a~well-ordered pair of sets $(S,T)$, the edge path in $\gn$
which is shown on Figure~\ref{fig:pst1} connects the vertices
$\fatb_S$ and $\fatb_T$. We denote this path by $p_{S,T}$ and call it
the {\bf standard path associated to} the well-ordered pair $(S,T)$.
\end{df}

We shall say that two well-ordered pairs of sets $(S,T)$ and $(S',T')$
are {\it disjoint} if $\{S,T\}\cap\{S',T'\}=\emptyset$. Clearly, two
well-ordered pairs of sets are disjoint if and only if the
corresponding paths $p_{S,T}$ and $p_{S',T'}$ have no endpoints in
common. The following theorem shows that a~much stronger statement is
true.

\begin{thm}\label{thm:pst}
Assume $(S,T)$ and $(S',T')$ are disjoint well-ordered pairs of sets,
which are not nested, then the corresponding standard paths $p_{S,T}$
and $p_{S',T'}$ are disjoint.
\end{thm}

\pr The informal idea of the proof is that we want to see that one of
the endpoints of the standard path is detectable from any vertex on
the path. Hence, roughly speaking, if two standard paths have a vertex
in common, then they would have to have an endpoint in common.

To start with, we define an operation $ds(-)$. Given a~full
$[n]$-tuple $\sigma=A_1\s\dots\s A_t$, pick the indices $1\leq
i_1<\dots<i_k\leq t$, such that $|A_j|\geq 2$ if and only if
$j\in\{i_1,\dots,i_k\}$. We now set
\[\dd(\sigma):=A_1\cup\dots\cup A_{i_1}\s A_{i_1+1}\cup\dots\cup A_{i_2}
\s\dots\s A_{i_{k-1}+1}\cup\dots\cup A_{i_k}\s A_{i_k+1}\cup\dots\cup
A_t.\] In the degenerate case $|A_1|=\dots=|A_t|=1$, we set
$\dd(\sigma):=[n]$.
 
Clearly, $\dd(\sigma)$ is again a~full $[n]$-tuple, which either
does not have any singletons, or has exactly one singleton as the last
set.  In general, $\sigma=\dd(\sigma)$ if and only if $\sigma$ either
does not have any singletons, or its last set is the only singleton.
 
Let us now consider a~well-ordered pair of sets $(S,T)$. As the first
case we assume that $|S|\geq 2$ and $|T|\geq |S|+2$. We now apply
$\dd(-)$ to the vertices of the standard path $p_{S,T}$. The obtained
full $[n]$-tuples are: $(S\s [n]\sm S)$, $(T\s [n]\sm T)$, and $(S\s
T\sm S\cup [n]\sm T)$.  In all cases, the first set in that full
$[n]$-tuple is indexing one of the endpoints of $p_{S,T}$, thus if
$p_{S,T}$ and $p_{S',T'}$ have a~vertex in common, then they also have
one of the endpoints in common.

In the remaining cases we still get the same possible patterns for
$\dd(\sigma)$ with one additional pattern: $\dd(\sigma)=[n]$. We get
this pattern in two cases:
\begin{itemize}
\item when $s=1$ and $\sigma=x_1\s\dots\s x_k\s
  x_{k+1},\dots,x_t,y_1,\dots,y_{n-t}$ with some $1\leq k\leq t$;
\item when $t=s+1$ and $\sigma=x_1\s\dots\s x_s\s x_{s+1}\s
  y_1,\dots,y_{n-t}$.
\end{itemize}
In other words, given $\sigma$ from $p_{S,T}$ we can always determine
either $S$ or $T$, except for one case. In this case, we have
$\sigma=a_1\s\dots\s a_k\s b_1,\dots,b_{n-k}$. There are two
possibilities for the well-ordered pair of sets $(S,T)$. Either $S=\{a_1\}$
and $\{a_1\,\dots,a_k\}\subseteq T$, or $S=\{a_1,\dots,a_{k-1}\}$ and
$T=\{a_1,\dots,a_k\}$.

Assume now that the paths $p_{S,T}$ and $p_{S',T'}$ do intersect.
Since $(S,T)$ and $(S',T')$ are disjoint, the paths must intersect at
an internal point. By what is said above we can assume without loss of
generality that $S=\{a_1\}$, $\{a_1,\dots,a_k\}\subseteq T$,
$S'=\{a_1,\dots,a_{k-1}\}$, and $T'=\{a_1,\dots,a_k\}$. However, this
means that the pairs $(S,T)$ and $(S',T')$ are nested, contradicting
our assumptions.  \qed


\section{From comparable matchings to symmetry breaking labelings}\label{sect:9}

\subsection{Perfect matchings induce symmetry breaking labelings}

\begin{thm}\label{thm:a} {\bf (Theorem A).} $\,$

\nin Let $n$ be an arbitrary natural number, and let
$\lambda:\cn_n^2\ra\{0,1\}$ be a~blank and compliant binary labeling
on the $n$-nodes of second level. Assume that there exists a perfect
matching on the graph $\cm_\lambda$, then there exists a~symmetry
breaking labeling on the $n$-nodes of third level.
\end{thm}
\pr Let $\mu$ denote the perfect matching on the graph
$\cm_\lambda$. We now proceed to give a~rule defining a binary node
labeling $\rho:\cn_n^3\ra\{0,1\}$ on the $n$-nodes of the third level.
Take $v\in\cn_n^3$, $v=v_1\ds v_2\ds (S,x)$. To start with we set
\[\rho^\defa(v):=\lambda(\parent(v)),\]
and call this a {\it default value} of $\rho$. The rule for defining
the value of $\rho$ distinguishes 3 cases.

\vskip5pt

\nin {\bf Case 1.} {\it Assume $|S|\leq n-2$}. In this case, we set
$\rho(v):=\rho^\defa(v)$.

\vskip5pt

\nin {\bf Case 2.} {\it Assume $|S|=n-1$}. In this case, there exists
$y\in [n]$, such that $S=[n]\sm y$. Since $\carr(v_2)\supseteq S$, we
have $|\carr(v_2)|\geq n-1$. If $|\carr(v_2)|=n-1$, then we set
$\rho(v):=\rho^\defa(v)$.  Else, we must have $\carr(v_2)=[n]$. Since
$\col(v_2)=S$, we see that $v_2$ is an edge in $\gn$. At the same time
$v_1$ is a full $[n]$-tuple, so $v_1\ds v_2$ is an edge in $\gn^2$. If
the vertices of $\gn^2$ connected by this edge are matched under
$\mu$, then we set $\rho(v):=\rho^\defa(v)$, else we set $\rho(v):=1$.

\vskip5pt

\nin {\bf Case 3.} {\it Assume $|S|=n$}. In other words, we have
$S=[n]$. In this case $\alpha=v_1\ds v_2$ is a~vertex of $\gn^2$. If
this vertex is not monochromatic with respect to $\lambda$, then we
set $\rho(v):=\rho^\defa(v)$. If $\alpha$ is monochromatic, then we
know that it has been matched, since we assumed that the matching
$\mu$ is perfect. In particular, the label $\id_\mu(\alpha)$ is
well-defined.  We now complete our definition of $\rho$ by setting
\[\rho(v):=\begin{cases}
0, &\text{ if }x=\id_\mu(\alpha);\\
1, &\text{ if }x\neq\id_\mu(\alpha).
\end{cases}\]

The value $\rho(v)$ has now been defined for all $v\in\cn_n^3$, and we
would like to summarize by saying that $\rho(v)$ may be different from
the default value $\rho^\defa(v)$ only in the following two cases:
\begin{itemize}
\item if $|S|=n-1$ and $v_1\ds v_2$ is a~matching edge; 
\item if $S=[n]$, $v_1\ds v_2$ is a~monochromatic vertex, and
  $x\neq\id_\mu(v_1\ds v_2)$.
\end{itemize}

To see that the node labeling $\rho$ is symmetry breaking, we need to
verify that it is compliant, and that it does not have any
monochromatic vertices. We start with proving that $\rho$ is
compliant.  Assume, we have two boundary nodes $v,w\in\cn_n^3$,
$v=v_1\ds v_2 \ds (S,x)$, $w=w_1\ds w_2\ds (T,y)$, such that
$|\supp(v)|=|\supp(w)|\leq n-1$.  By definition of the carrier this
means that $|\carr(v)|=|\carr(w)|$.  Let $\varphi:\supp(v)\ra\supp(w)$
be the unique order-preserving bijection. Assume furthermore that
$\varphi(v)=w$.  Specifically, this means that $w_1=\varphi(v_1)$,
$w_2=\varphi(v_2)$, $T=\varphi(S)$, and $y=\varphi(x)$.

We now show that under these conditions, we have
$\rho(v)=\rho^\defa(v)$ and $\rho(w)=\rho^\defa(w)$. First, since
$T=\varphi(S)$, we have $|S|=|T|$. Second, we have
$S\subseteq\supp(v)$, so $|S|\leq |\supp(v)|\leq n-1$. If $|S|=|T|\leq
n-2$, then $\rho(v)=\rho^\defa(v)$ and $\rho(w)=\rho^\defa(w)$ by the
Case~1 of our rule. If, on the other hand, $|S|=|T|=n-1$, then
$|\supp(v)|=|\supp(w)|=n-1$, and this time $\rho(v)=\rho^\defa(v)$ and
$\rho(w)=\rho^\defa(w)$ by the Case 2 of our rule for defining~$\rho$.

Next, let us show that $\varphi(\parent(v))=\parent(w)$. By the
calculation after Definition~\ref{df:parent}, we have
$\parent(v)=\gamma_1\ds (\wti S,x)$, where $(\wti S,x)=v_2\dar x$, and
$\gamma_1=v_1\dar\wti S$. Clearly, the operation $\dar$ commutes with
$\varphi$, so we have
\[\varphi(\wti S,x)=\varphi(v_2\dar x)=\varphi(v_2)\dar\varphi(x)=w_2\dar y\]
and
\[\varphi(\gamma_1)=\varphi(v_1\dar\wti S)=\varphi(v_1)\dar\varphi(\wti S)=
w_1\dar (w_2\dar y),\]
so 
\[\varphi(\gamma_1\ds(\wti S,x))=\varphi(\gamma_1)\ds(\varphi(\wti S),\varphi(x)))
=\parent(w).\]

Let us now verify that $\rho$ has no monochromatic vertices. Let
$\sigma=\sigma_1\ds\sigma_2\ds\sigma_3$ be an arbitrary vertex of $\gn^3$.
We consider two cases.

\vskip5pt

\nin {\bf Case 1.} {\it Assume the vertex $\sigma_1\ds\sigma_2$ is not
  monochromatic.}  We show that $\rho(v)=\rho^\defa(v)$ whenever $v$
is a~node of the third level adjacent to the vertex $\sigma_1 \ds
\sigma_2\ds\sigma_3$. Assume $\sigma_3=A_1\s\dots\s A_t$. If $x\in
A_k$, such that $|A_1\cup\dots\cup A_k|\leq n-2$, then $\rho(v)=
\rho^\defa(v)$ by definition. If, on the other hand, $x\in A_t$, then
$\sigma\dar x=\sigma_1\ds\sigma_2\ds([n],x)$. Since the vertex
$\sigma_1 \ds\sigma_2$ is not monochromatic, we again get
$\rho(v)=\rho^\defa(v)$. The last remaining case is when $x\in A_k$,
such that $|A_1\cup\dots\cup A_k|=n-1$. This is only possible if
$\sigma_3=A_1\s\dots\s A_{t-1}\s y$, and $x\in A_{t-1}$. If that
happens we have $\sigma\dar x=\tau_1\ds\tau_2\ds ([n]\sm y,x)$. If now
$\tau_1\ds\tau_2$ is not an edge, we must have $\rho(v)=
\rho^\defa(v)$.  Finally, if $\tau_1\ds\tau_2$ is an edge, then
$\sigma_1\ds\sigma_2$ is one of its endpoints. However, the vertex
$\sigma_1\ds\sigma_2$ is not monochromatic by our assumption, and so
$\tau_1\ds\tau_2$ cannot be a~matching edge; hence again we get
$\rho(v)= \rho^\defa(v)$.

We have now proved that $\rho(v)=\rho^\defa(v)$ whenever $v$ is a~node
of the third level adjacent to the vertex $\sigma_1\ds\sigma_2 \ds
\sigma_3$. On the other hand, the set of parents of these nodes is
precisely the set of nodes of the second level adjacent to $\sigma_1
\ds \sigma_2$, because $\parent(\sigma\dar x)= (\sigma_1\ds \sigma_2)
\dar x$. Since we assumed that the vertex $\sigma_1\ds \sigma_2$ is
not monochromatic, we conclude that neither is the vertex~$\sigma$.

\vskip5pt

\nin {\bf Case 2.} {\it Assume the vertex $\sigma_1\ds\sigma_2$ is
  monochromatic.} Since $\mu$ is a~perfect matching, there exists
an~edge in $\gn^2$ matching $\sigma_1\ds\sigma_2$ to some other
monochromatic vertex. Set $c$ to be the label of that edge, and assume
$\sigma_3=A_1\s\dots\s A_t$. We now distinguish three further
subcases.

\vskip5pt

\nin {\bf Case 2a.}  Assume $|A_t|\geq 2$. If $x\in A_t$, then set
$v_r:=\sigma\dar x=\sigma_1\ds\sigma_2\ds ([n],x)$. This is the node
with color $x$ adjacent to $\sigma$. By our rule, if $x\neq c$, then
$\rho(v_r)=1$. Since $|A_t|\geq 2$, there exists $x\in A_t$ such that
$x\neq c$, so at least one of the nodes adjacent to $\sigma$ has
label~$1$. On the other hand, let $v_c$ be the node of color~$c$
adjacent to $\sigma$. If $c\in A_t$, then this node has label~$0$.
Otherwise, we have $c\in A_k$, for some $1\leq k<t$. In this case, we
have $v_c:=\sigma\dar c=\ti\sigma_1\ds\ti\sigma_2\ds (A_1\cup\dots\cup
A_k,c)$. Since $|A_1\cup\dots\cup A_k|\leq n-2$, we get $\rho(v_c)=0$
again. Either way, we have nodes with different labels adjacent to
$\sigma$, so $\sigma$ is not monochromatic.

\vskip5pt

\nin {\bf Case 2b.}  Assume $|A_t|=1$, $A_t=\{y\}$, with $y\neq c$. As
above we calculate $\rho(v_y)=1$. Take now $x\in A_{t-1}$. If
$\rho(x)=\rho^\defa(x)$, then $\rho(x)=0$, since the vertex
$\sigma_1\ds\sigma_2$ is $0$-monochromatic. Otherwise, we have
$\sigma\dar x=\ti\sigma_1\ds\ti\sigma_2\ds([n]\sm y,x)$, and
$\ti\sigma_1\ds\ti\sigma_2$ is a~matching edge. This is impossible,
since that edge would be labeled $y$, and have $\sigma_1\ds\sigma_2$
as one of its endpoints, contradicting the assumption~$y\neq c$.
In either case we have two nodes adjacent to $\sigma$ with different
values of~$\rho$, so $\sigma$ is not monochromatic.

\vskip5pt

\nin {\bf Case 2c.}  Assume $|A_t|=1$, $A_t=\{c\}$. By the calculation
in Case 2a, we get $\rho(v_c)=0$. Take any $x\in A_{t-1}$, so
$v_x=\sigma\dar x=\ti\sigma_1\ds\ti\sigma_2\ds([n]\sm c,x)$. Now
$\ti\sigma_1\ds\ti\sigma_2$ is an edge labeled $c$ which has
$\sigma_1\ds\sigma_2$ as an endpoint. By assumptions above this means
that $\ti\sigma_1\ds\ti\sigma_2$ is a~matching edge, and so we get
$\rho(v_x)=1$. Again, we have different values of $\rho$ on the nodes
adjacent to $\sigma$, so $\sigma$ is not monochromatic.

\vskip5pt

\noindent
This finishes the proof of the theorem.
\qed

\begin{crl}\label{crl:a}
 Let $n$ be an arbitrary natural number, and let $\cb$ be an~arbitrary
 set of patterns in $[n]$. If there exists a~perfect matching on the
 graph $\cm_{\lambda_\cb}$, then there exists a~symmetry breaking
 labeling on the $n$-nodes of third level.
\end{crl}
\pr The binary node labeling $\lambda_\cb$ associated to the set of
patterns $\cb$ is always blank and compliant, see
Proposition~\ref{prop:bcl}, hence the statement follows from
Theorem~\ref{thm:a}. \qed

\subsection{Non-intersecting path systems induce perfect matchings} $\,$


\begin{df}
Let $n$ be an arbitrary natural number, $n\geq 2$. The linear
Diophantine equation in $n-1$ variables
\begin{equation}
x_1\binom{n}{1}+x_2\binom{n}{2}+\dots+x_{n-1}\binom{n}{n-1}=1
\end{equation}
is called {\bf binomial Diophantine} equation associated to $n$.
\end{df}

A solution $(x_1,\dots,x_{n-1})$ to the binomial Diophantine equation
associated to $n$ is called {\it primitive} if $x_1=1$,
$x_2\in\{0,1\}$, and $x_i\in\{-1,0,1\}$, for all $i=3,\dots,n-1$. For
example $(1,1,-1,0,0)$ is a~primitive solution for $n=6$, and
$(1,0,-1,1,0,0,0,0,-1,-1,0)$ is a~primitive solution for $n=12$.

We shall now consider families of proper subsets of the set $[n]$,
i.e., $\Sigma\subseteq 2^{[n]}\sm\{\emptyset,[n]\}$, which we call
{\it proper} families. Let $C^n_t$ be the family of all subsets of
$[n]$ of cardinality $t$. It is a~proper family if $1\leq t\leq n-1$.
 
\begin{df}
A proper family $\Sigma\subseteq 2^{[n]}$ is called {\bf cardinal} if
the following is satisfied: whenever $S\in\Sigma$, we have
$C^n_{|S|}\subseteq\Sigma$. In other words, if $\Sigma$ contains one
set with $k$ elements, then it contains all sets with $k$ elements,
which are subsets of~$[n]$.
\end{df} 

A cardinal family $\Sigma$ can be described simply by specifying the
list of cardinalities $C(\Sigma)\subseteq\{1,\dots,n-1\}$ of the sets
in~$\Sigma$, namely $\Sigma=\cup_{t\in C(\Sigma)}C^n_t$. Two proper
families $\Sigma$ and $\Lambda$ are {\it disjoint} if and only if they
are disjoint as sets. Clearly, cardinal families $\Sigma$ and
$\Lambda$ are disjoint if and only if the corresponding cardinality
sets $C(\Sigma)$ and $C(\Lambda)$ are disjoint.

Assume now that we are given $n\geq 2$, and that
$\fatx=(x_1,\dots,x_{n-1})$ is a~primitive solution to the binomial
Diophantine equation associated to $n$. Set
$I_\fatx:=\{i\,|\,i\in[n],\,x_i=1\}$ and
$J_\fatx:=\{j\,|\,j\in[n],\,x_j=-1\}$, in particular $1\in
I_\fatx$. Furthermore, set $\Sigma_\fatx:=\{S\,|\,S\subset[n],\,|S|\in
I_\fatx\}$ and $\Lambda_\fatx:=\{T\,|\,T\subset[n],\,|T|\in
J_\fatx\}$. Clearly, $\Sigma_\fatx$ and $\Lambda_\fatx$ are proper
families of subsets. They are disjoint because $I_\fatx$ and $J_\fatx$
are disjoint. Finally, since $\fatx$ is a~primitive solution to the
binomial Diophantine equation associated to $n$ we have
$|\Sigma_\fatx|=|\Lambda_\fatx|+1$. We say that the proper families of
subsets $\Sigma_\fatx$ and $\Lambda_\fatx$ are {\it associated}
to~$\fatx$.

\begin{df}
Assume we are given two proper set families $\Sigma$ and $\Lambda$,
such that $|\Sigma|=|\Lambda|$. A~{\bf non-intersecting path system}
between $\Sigma$ and $\Lambda$ consists of a~bijection
$\varphi:\Sigma\ra\Lambda$ together with a~set of disjoint edge paths
$\{q_{S,\varphi(S)}\}_{S\in\Sigma}$, such that each path
$q_{S,\varphi(S)}$ connects $\fatb_S$ with~$\fatb_{\varphi(S)}$.
\end{df}

\begin{thm}\label{thm:pathsol}{\bf (Theorem B).}

\nin Assume $\fatx=(x_1,\dots,x_{n-1})$ is a~primitive solution to the
binomial Diophantine equation associated to some $n\geq 2$, and
$\Sigma_\fatx$ and $\Lambda_\fatx$ are the associated proper families
of subsets of $[n]$. If there exists a non-intersecting path system
between $\Sigma_\fatx\sm\{n\}$ and $\Lambda_\fatx$, then there exists
a~compliant symmetry breaking binary labeling on the $n$-nodes of the
third level.
\end{thm}
\pr Set as above $I_\fatx:=C(\Sigma_\fatx)$ and
$J_\fatx:=C(\Lambda_\fatx)$. We have a~bijection
$\varphi:\Sigma_\fatx\sm\{n\}\ra\Lambda_\fatx$, and a~family of
disjoint edge paths in $\gn$,
$\{q_{S,\varphi(S)}\}_{S\in\Sigma_\fatx\sm\{n\}}$, such that each path
$q_{S,\varphi(S)}$ connects $\fatb_S$ with~$\fatb_{\varphi(S)}$.  Let
$\cb_\fatx=(\cb_1,\dots,\cb_{n-1})$ be the set of patterns associated
to vector $\fatx$, and consider the associated node labeling
$\lambda=\lambda_{\cb_\fatx}:\cn_n^2\ra\{0,1\}$.

We have a bipartite graph $\cm_\lambda$, and we are looking for
a~perfect matching on this graph. This graph consists of subgraphs
$\cm_\lambda(\sigma)$ where $\sigma=A_1\s\dots\s A_t$ ranges through
vertices of $\gn$, We can start by taking some matchings on these
subgraphs and then eliminating the remaining critical vertices.
Theorem~\ref{thm:pat} describes $\cm_\lambda(\sigma)$ combinatorially,
for each~$\sigma$.  All these graphs are isomorphic to
$\gn(\Omega(\cb_\fatx,\sigma),V)$.  By Remark~\ref{rm:fp}, the set
$\Omega(\cb_\fatx,\sigma)$ does not contain any full $V$-tuples
whenever $t\geq 3$, so in these cases $\cm_\lambda(\sigma)$ will have
a~perfect matching: we can take the standard matching with respect to
any order on $[n]\sm V$ and then apply
Theorem~\ref{thm:main_match}(3).

When $t=2$, we have $\sigma=S\s[n]\sm S=\fatb_S$, say
$S=\{x_1,\dots,x_k\}$, for $x_1<\dots<x_k$. In this case the standard
matchings on $\cm_\lambda(\sigma)$ are not perfect. They have
critical vertices which are in a~bijective correspondence with full
$V$-prefixes. Going back to the definition of the set of patterns
associated to $\fatx$, we distinguish 3 cases.
\begin{enumerate}
\item[{\bf Case 1.}] If $|S|\in I_\fatx$, then the set
  $\Omega(\cb_\fatx,\sigma)$ has a~unique full $V$-prefix, namely
  $x_1\s\dots\s x_k$. Thus for any order on $[n]\sm V$ the standard
  matching will have a~unique critical vertex.

\item[{\bf Case 2.}] If $|S|\in J_\fatx$, then the set
  $\Omega(\cb_\fatx,\sigma)$ has 3 full $V$-prefixes, namely
  $x_1\s\dots\s x_k$, $x_1,x_2\s x_3\s\dots\s x_k$, and $x_1\s
  x_2,x_3\s \dots\s x_k$. Note, that in this case we must have $k\geq
  3$. Thus for any order on $[n]\sm V$ the standard matching will have
  the corresponding 3 critical vertices.

\item[{\bf Case 3.}] If $|S|\notin I_\fatx\cup J_\fatx$, then
  $\Omega(\cb_\fatx,\sigma)$ is empty, and it follows from
  Theorem~\ref{thm:main_match}(2) that the standard matching is
  perfect.
\end{enumerate}

Finally, when $t=1$, we have $\sigma=[n]$. In this case
Theorem~\ref{thm:main_match}(1) applies and we have a~unique critical
vertex which depends on the chosen order.

We now use conductivity in flip graphs, as developed in
Section~\ref{sect:5}, to find edge paths in $\gn^2$ connecting all the
critical vertices. Let us fix $S\in\Sigma_\fatx\sm\{n\}$, and take the
corresponding path $q_{S,\varphi(S)}$. We shall be traversing that
path starting from $\fatb_{\varphi(S)}$ and going towards $\fatb_S$,
so we let $w_1:=\fatb_{\varphi(S)},w_2,\dots,w_{d-1},w_d:= \fatb_S$
denote the vertices on the path listed in that order. Note that $d$
must be odd. For $k=1,\dots,d-1$, let $y_k$ be the label of the edge
between $w_k$ and $w_{k+1}$. By Lemma~\ref{lm:sc-} one can choose the
order $R$, so that the standard matching can be extended to match two
of the critical vertices to each other, and there will exist
a~semi-augmenting edge path connecting the remaining critical vertex
to some $y_1$-connector of the second type~$\tau_1^f$, which is proper
with respect to~$\fatb(\varphi(S))$.

Let $\tau_2^s$ be the unique vertex connected to $\tau_1^f$ by the
edge with label $y$.  Clearly, $\tau_2^s$ is a~$y_1$-connector of the
second type, which by the identity~\eqref{eq:vf} is proper with
respect to~$w_2$. By Lemma~\ref{lm:c1}, there exists an~order $R$ and
a~non-augmenting edge path in the graph $\cm_\lambda(w_2)$ with
respect to $\mu_R$, connecting $\tau_2^s$ to some $y_2$-connector of
the first type $\tau_2^f$, which is proper with respect to~$w_2$. We
now let $\tau_3^s$ be the unique vertex connected to $\tau_2^f$ by the
edge with the label $y_2$, which is a~$y_2$-connector of the first
type proper with respect to~$w_3$. We then repeat that argument for
the graph $\cm_\lambda(w_3)$, using Lemma~\ref{lm:c2} instead.

Eventually, we will arrive at a vertex $\tau_d^s$ in
$\cm_\lambda(\fatb_S)$. Since $d$ is odd, $\tau_d^s$ is
a~$y_{d-1}$-connector of the first type, and it is proper with respect
to~$\fatb_S$. We now employ Lemma~\ref{lm:sc+}, which tells us that
there exists an order $R$ and a~semi-augmenting path with respect to
$\mu_R$ which connects $\tau_d^s$ with the unique critical vertex in
$\cm_\lambda(\fatb_S)$.  Concatenating all these paths will yield an
augmenting path which connects the two critical vertices in
$\cm_\lambda(\fatb_{\varphi(S)})$ and $\cm_\lambda(\fatb_S)$. Applying
the transformation from Definition~\ref{df:mdef} to that path will
yield a new matching, where these two critical vertices are now
matched. Doing this for all paths $q_{S,\varphi(S)}$, when $S$ ranges
over all subsets in $\Sigma_\fatx\sm\{n\}$ will yield a~matching on
$\cm_\lambda$, with two of the critical vertices remaining: one in
$\cm_\lambda(n\s[n-1])$, and one in $\cm_\lambda([n])$. 

Note, that by Theorem~\ref{thm:pat}, we have
$\cm_\lambda([n])\cong\gn$ and $\cm_\lambda(n\s [n-1]) \cong
\gn(\Omega,n)$, where $\Omega=\{n\}$. To start with, consider the
standard matching in $\cm_\lambda([n])$ with respect to the order
$(1,\dots,n)$, by Theorem~\ref{thm:main_match}, we have a~unique
critical vertex $v$ indexed by $[n]\ds 1\s 2\s\dots\s n$. Let $w$ be
the vertex of $\cm_\lambda(n\s [n-1])$ indexed by $n\s[n-1]\ds 1\s
2\s\dots\s n$. Clearly, the vertices $v$ and $w$ are connected in
$\cm_\lambda$ by an edge labeled~$n$. By Lemma~\ref{lm:sc+}, there
exists an order on $[n-1]$, such that there exists a~semi-augmenting
path connecting the unique critical vertex with $w$. In fact the order
$(1,\dots,n-1)$ will do, in which case the unique critical vertex will
be $n\s[n-1]\ds n\s 1\s\dots\s n-1$, and the semi-augmenting path can
be given explicitly: $n\s 1\s\dots\s n-1\ra 1,n\s 2\s\dots\s n-1\ra
1\s n\s 2\s\dots\s n-1\ra\dots\ra 1\s 2\s\dots\s n-1,n\ra
1\s 2\s\dots\s n$. Concatenating this path with the edge between $v$
and $w$ yields an~augmenting path which eliminates the last two
critical vertices, resulting in a~perfect matching on~$\cm_\lambda$.
\qed

\subsection{Comparable matchings induce non-intersection path systems} $\,$

\nin
Producing a~non-intersecting path system for $n=6$, and
$\fatx=(1,1,-1,0,0)$ has been done by hand in \cite{wsb6}.
Unfortunately, doing it directly appears prohibitive for larger values
of~$n$. We now look for further structures which will help us
construct non-intersecting path systems.

\begin{df}
A {\bf comparable matching} between disjoint proper families $\Sigma$
and $\Lambda$ is a~bijection $\varphi:\Sigma\ra\Lambda$, such that for
any $S\in\Sigma$, either $(S,\varphi(S))$ or $(\varphi(S),S)$ is
a~well-ordered pair. We say that this well-ordered pair is {\bf
  associated} to $S$. 

The comparable matching $\varphi$ is called {\bf non-nested} if for
any $S,T\in\Sigma$ the associated well-ordered pairs are not nested.
\end{df}

Given disjoint proper families $\Sigma$ and $\Lambda$, the set of
comparable matchings $\varphi:\Sigma\ra\Lambda$ can be partially
ordered as follows. Assume we have two subsets $S,T\subseteq[n]$, such
that one contains the other one. If $S\subset T$, then we set
$l(S,T):=|T\sm S|$, else set $l(S,T):=|S\sm T|$; this is a~distance
between $S$ and~$T$. Let $L_\varphi$ denote the multiset of distances
$\{l(S,\varphi(S))\,|\,S\in\Sigma\}$. We define an~associated function
$\dist_\varphi$ on the set of natural numbers, by setting
$\dist_\varphi(d)$ to be the number of occurrences of $d$ in
$L_\varphi$. Since $L_\varphi$ is a~finite multiset, the value
$\dist_\varphi(d)$ is different from $0$ for only finitely many values
of~$d$. 

Assume now we are given two comparable matchings
$\varphi,\psi:\Sigma\ra\Lambda$. If the functions $\dist_\varphi$ and
$\dist_\psi$ are identical, we say that $\varphi$ and $\psi$ are
incomparable. Otherwise, let $k$ be the maximal index such that
$\dist_\varphi(d)\neq\dist_\psi(d)$. We now say that
$\varphi\prec\psi$ if $\dist_\varphi(d)<\dist_\psi(d)$, and we say
that $\varphi\succ\psi$ if $\dist_\varphi(d)>\dist_\psi(d)$. Clearly,
this is a~well-defined partial order on the set of all comparable
matchings between $\Sigma$ and $\Lambda$, which we call {\it
  distance-lexicographic} order.

\begin{prop} \label{pr:ncm}
Assume we have proper families $\Sigma$ and $\Lambda$, and
a~comparable matching $\varphi:\Sigma\ra\Lambda$, then there exists
a~non-nested comparable matching $\psi:\Sigma\ra\Lambda$.
\end{prop}

\pr Without loss of generality, we can assume that $\varphi$ is chosen
to be a~comparable matching which is minimal with respect to the
distance-lexicographic order defined above. If $\varphi$ is
non-nested, then we are done, so assume this is not the case and take
any pair $S,T\in\Sigma$ such that the associated well-ordered pairs
are nested. Without loss of generality, swapping $\Sigma$ and
$\Lambda$ if necessary, we can assume that $S\subset\varphi(S)$,
$S\subset T$, and $S\subset\varphi(T)$. We then have two cases, either
we have $S\subset T\subset \varphi(T)\subset \varphi(S)$, or
$S\subset\varphi(T)\subset T \subset\varphi(S)$.

Define a~new bijection $\psi:\Sigma\ra\Lambda$ as follows:
$\psi(A):=\varphi(A)$, for $A\neq S,T$, $\psi(S):=\varphi(T)$, and
$\psi(T):=\varphi(S)$. Clearly, $\psi$ is again a~comparable matching,
which precedes $\varphi$ in the distance-lexicographic order. This
contradicts the choice of~$\varphi$.  \qed

\begin{thm}\label{thm:main} {\bf (Theorem C).}

\nin Assume $\fatx=(x_1,\dots,x_{n-1})$ is a~primitive solution to the
binomial Diophantine equation associated to some $n\geq 2$, and
$\Sigma_\fatx$ and $\Lambda_\fatx$ are the associated proper families
of subsets of $[n]$. If there exists a~comparable matching between
$\Sigma_\fatx\sm\{n\}$ and $\Lambda_\fatx$, then there exists
a~compliant symmetry breaking binary labeling on the $n$-nodes of the
third level.
\end{thm}

\pr Consider a~comparable matching
$\varphi:\Sigma_\fatx\sm\{n\}\ra\Lambda_\fatx$. By
Proposition~\ref{pr:ncm} we might as well assume that $\varphi$ is
non-nested. By Theorem~\ref{thm:pst} the family
$\{p_{S,\varphi(S)}\}_{S\in\Sigma_\fatx\sm\{n\}}$ is
a~non-intersecting path system, so the result follows from
Theorem~\ref{thm:pathsol}.  \qed

\begin{dcc}
Theorem~\ref{thm:main} means that in the standard computational model
the existence of a~comparable matching between disjoint cardinal
proper families of subsets of $[n]$ implies the existence of
a~wait-free protocol solving Weak Symmetry Breaking in $3$ rounds.
\end{dcc}


\section{New upper bounds for $\msb(n)$}\label{sect:10}

\subsection{The formulation of the main theorem and some set theory notations} $\,$

\nin Our goal now is to use Theorem~\ref{thm:main} to improve upper
bounds for the symmetry breaking function $\msb(n)$. Our most definite
result is to show that there are infinitely many values of $n$ for
which $\msb(n)\leq 3$.


We now return to considering Theorem~\ref{thm:6n}.  The case $t=1$ has
been previously settled in \cite{wsb6}. To deal with the case $t=2$,
we need to start with an~appropriate Diophantine equation. It just so happens
that we have the arithmetic identity:
\begin{equation}
\label{eq:507}
\binom{12}{1}+\binom{12}{4}=\binom{12}{0}+
\binom{12}{3}+\binom{12}{9}+\binom{12}{10}.
\end{equation}
Indeed, both sides of \eqref{eq:507} are equal to $507$. That
particular number has only technical significance - it shows the
number of augmenting paths which we will need to fix in the initial
standard matching, in order to arrive at a~perfect matching. Once we
have the identity~\eqref{eq:507}, we can use computer search to show
the existence of a~comparable matching between the corresponding
disjoint cardinal proper families of subsets of~$[12]$. Clearly, this
approach will only work for small values of $n$, and to deal with the
general case, we need to move beyond the direct computer search.

Before proceeding with the proof of Theorem~\ref{thm:6n} we need
a~little bit of terminology.  We shall think about subsets of the set
$[n]$ in terms of their support vector, i.e., we identify a~subset
$S\subseteq [n]$ with an $n$-tuple $\chi_S=(a_1,\dots,a_n)$, where
$a_i=1$ if $i\in S$ and $a_i=0$ otherwise.

\begin{df}
Let $\alpha$ be any tuple of length at most $n$, consisting of $0$'s
and $1$'s. We let $\suf{\alpha}_n$ denote the set of all subsets $S$
whose support vector ends with $\alpha$. If $n$ does not matter, we
shall drop it, and simply write $\suf\alpha$.
\end{df}

So, if $\alpha=(a_1,\dots,a_k)$, then the set $\suf\alpha$ consists of
all subsets~$S$, for which we have $\chi_S=(b_1,\dots,b_{n-k},\ab
a_1,\dots,a_k)$, or, in other words, for $i=n-k+1,\dots n$, we have
$i\in S$ if and only if $a_{i+k-n}=1$. If $k=0$, i.e., $\alpha$ is an
empty tuple, we have $\suf\alpha_n=2^{[n]}$, consistently with the
standard notation. As another example $\suf{(0,1)}_n$ denotes the set
of all subsets $S$ such that $n\in S$, but $n-1\notin S$.

We shall use the short-hand notation skipping the commas and the
brackets, and write $\suf{01}$ instead of $\suf{(0,1)}$. Furthermore,
we shall use the square brackets to encode the repetitions: when
$\alpha$ is any tuple of $0$'s and $1$'s, $[\alpha]^k$ denotes the
tuple obtained by repeating $\alpha$ $k$ times. For example, $0[01]^3$
stands for $0010101$.  We shall also use the notation $[\alpha]^*$ to
say that $\alpha$ is repeated a~certain number of times, without
specifying the number of repetitions, which can also be~$0$. For
example,
\[\suf{0[01]^*}=\suf 0\cup\suf{001}\cup\suf{00101}\cup\dots,\]
and we have
$\suf{0[01]^*}_4=\{0000,0010,0100,0110,1000,1010,1100,1110,0001,1001\}$,
while
$\suf{1[10]^*}_4=\{0001,0011,0101,0111,1001,1011,1101,1111,0110,1110\}$.

We let $[\alpha]^\infty_{\leq n}$ denote the tuple obtained by first
repeating $\alpha$ infinitely many times, and then truncating it at
position~$n$; it is as much of repeated $\alpha$ as is possible to fit
in the first $n$ slots. For example, for $\alpha=01$, we get
$[01]^\infty_{\leq 4}=0101$ and $[01]^\infty_{\leq 3}=101$.

\subsection{Some useful set decompositions}$\,$

\nin In order to define a bijection, which is crucial for the proof of
Theorem~\ref{thm:6n}, we need a~number of specific set
decompositions. In the formulations below, we use the symbol
$\,\,\,\bar{}\,\,\,$ to denote negation, so $\bar 0=1$ and $\bar 1=0$.

\begin{lm}\label{lm:alpha}
Whenever $\alpha=(\alpha_1,\dots,\alpha_t)$ is a~tuple consisting of
$0$'s and $1$'s, where $n\geq t$, we have the following decomposition
into disjoint subsets:
\begin{equation}\label{eq:alpha}
2^{[n]}=[\alpha]^\infty_{\leq n}\cup\bigcup_{i=1}^t
\suf{\bar\alpha_i\alpha_{i+1}\dots\alpha_t[\alpha]^*}.
\end{equation}
\end{lm}
\pr Take any $S\in 2^{[n]}$, and read its support vector
$\chi_S=(a_1,\dots,a_n)$ from right to left, starting with $a_n$. The
first position, where $\chi_S$ deviates from $[\alpha]^\infty_{\leq
  n}$ will show in which of the disjoint sets of the right hand side
of~\eqref{eq:alpha} the set $S$ lies. \qed

\begin{crl}\label{crl:exp}
For arbitrary $p\geq 1$, we have following identities:
\[2^{[2p]}=\suf{0[01]^*}\cup\suf{11[01]^*}\cup [01]^p=\suf{1[10]^*}\cup
\suf{00[10]^*}\cup [10]^p\]
\[2^{[2p+1]}=\suf{0[01]^*}\cup\suf{11[01]^*}\cup 1[01]^p=\suf{1[10]^*}\cup
\suf{00[10]^*}\cup 0[10]^p\]
\end{crl}
\pr Follows from Lemma~\ref{lm:alpha} by substituting $\alpha=01$ and
$\alpha=10$ into the equation~\eqref{eq:alpha} and considering the two
cases when $n$ is even or odd.  \qed

\begin{crl} \label{crl:001}
For any $p\geq 2$ we have the following identities, where all unions
on the right hand side are disjoint
\begin{multline}\label{eq:001even}
\suf{0[01]^*}_{2p}= \suf{10[01]^*}\cup\suf{1[10]^* 00[01]^*}
\cup\suf{00[10]^* 00[01]^*}\cup \\
\cup\{[10]^k00[01]^{p-k-1}\,|\,0\leq k\leq p-1\},
\end{multline}
\begin{multline}\label{eq:001odd}
\suf{0[01]^*}_{2p+1}=\suf{10[01]^*}\cup\suf{1[10]^* 00[01]^*}
\cup\suf{00[10]^* 00[01]^*}\cup 0[01]^p\cup\\
\cup\{0[10]^k00[01]^{p-k-1}\,|\,0\leq k\leq p-1\}.
\end{multline}
\end{crl}
\pr Throughout the proof all the unions will be disjoint. We start
with the identity
\begin{equation}\label{eq:c1}
\suf{0[01]^*}_{2p}=\suf{00[01]^*}_{2p}\cup\suf{10[01]^*}_{2p}.
\end{equation}
We expand the first term
\begin{equation}\label{eq:c2}
\suf{00[01]^*}_{2p}=\suf{00}_{2p}\cup\suf{0001}_{2p}\cup\dots\cup
\suf{00[01]^k}_{2p}\cup\dots\cup\suf{00[01]^{p-1}}_{2p}.
\end{equation}
By Corollary~\ref{crl:exp}, for all $0\leq k\leq p-2$ we get
\begin{equation}\label{eq:c3}
\suf{00[01]^k}_{2p}=\suf{1[10]^*00[01]^k}_{2p}\cup
\suf{00[10]^*00[01]^k}_{2p}\cup [10]^{p-k-1}00[01]^k,
\end{equation}
and, furthermore, we have $\suf{00[01]^{p-1}}_{2p}=00[01]^{p-1}$.
Taking the union of this equation with the equation~\eqref{eq:c3} for
all~$k$, we get the identity
\begin{multline}\label{eq:c4}
\suf{00[01]^*}_{2p}=\suf{1[10]^*00[01]^*}_{2p}\cup
\suf{00[10]^*00[01]^*}_{2p}\cup\\\cup
\{\suf{[10]^{p-k-1}00[01]^k}_{2p}\,|\,0\leq k\leq p-1\}.
\end{multline}
Substituting this into~\eqref{eq:c1} proves~\eqref{eq:001even}.

Proving the odd case \eqref{eq:001odd} needs a little modification.
The decomposition~\eqref{eq:c1} gets replaced with
\begin{equation}\label{eq:c1odd}
\suf{0[01]^*}_{2p+1}=\suf{00[01]^*}_{2p+1}\cup\suf{10[01]^*}_{2p+1}
\cup 0[01]^p,
\end{equation}
while the decomposition~\eqref{eq:c3} gets replaced with
\begin{equation}\label{eq:c3odd}
\suf{00[01]^k}_{2p+1}=\suf{1[10]^*00[01]^k}_{2p+1}\cup
\suf{00[10]^*00[01]^k}_{2p+1}\cup 0[10]^{p-k-1}00[01]^k,
\end{equation}
for $0\leq k\leq p-2$, and $\suff{00[01]^{p-1}}_{2p+1}=
\{000[01]^{p-1},100[01]^{p-1}\}$. Taking the union over all~$k$, and
then substituting back into~\eqref{eq:c1odd} will now
yield~\eqref{eq:001odd}.  \qed

\begin{prop}\label{prop:110}
We have the following identities, where all unions on the right hand
side are disjoint
\begin{multline}\label{eq:110even}
\suf{1[10]^*}_{2p}=\suf{11[01]^*}\cup\suf{1[10]^* 10[01]^*}
\cup\suf{00[10]^* 01[01]^*}\cup\\ \cup\{[10]^k [01]^{p-k}\,|\,0\leq
k\leq p-1\},\text{ for all }p\geq 3,
\end{multline}
\begin{multline}\label{eq:110odd}
\suf{1[10]^*}_{2p+1}= \suf{11[01]^*}\cup\suf{1[10]^* 10[01]^*}
\cup\suf{00[10]^* 01[01]^*}\cup 1[01]^p\cup \\
\cup\{0[10]^k[01]^{p-k}\,|\,0\leq k\leq p-1\},\text{ for all }p\geq 2.
\end{multline}
\end{prop}
\pr Throughout the proof all the unions will be disjoint.  We start
with proving~\eqref{eq:110even}. By definition of $[]^*$-notation, we
have the identity
\begin{equation}\label{eq:p0}
\suf{1[10]^*}_{2p}=\suf{1}_{2p}\cup\suf{1[10]^*10}_{2p}.
\end{equation} 
On the other hand, we have
$\suff{1}_{2p}=\suff{11}_{2p}\cup\suff{01}_{2p}$. Substituting this
in~\eqref{eq:p0} we arrive at
\begin{equation}\label{eq:p1}
\suf{1[10]^*}_{2p}=\suff{11}_{2p}\cup\suff{01}_{2p}\cup\suf{1[10]^*10}_{2p}.
\end{equation}
Next, we shall derive a~formula for the term $\suff{01}_{2p}$. We
start with the identity
\begin{equation}\label{eq:p1b}
2^{[2p-2]}=\suff{11[01]^*}_{2p-2}\cup\suff{0[01]^*}_{2p-2}\cup[01]^{p-1}
\end{equation}
which was proved in Corollary~\ref{crl:exp}. By definition of $[]^*$,
we have
\begin{equation}\label{eq:p1c}
\suff{0[01]^*}_{2p-2}=\suff{0}_{2p-2}\cup\suff{001}_{2p-2}\cup\dots
\cup\suff{0[01]^k}_{2p-2}\cup\dots\cup\suff{0[01]^{p-2}}_{2p-2}.
\end{equation}
Using Corollary~\ref{crl:exp} again, we derive the following identity
for all $0\leq k\leq p-3$ 
\begin{equation}\label{eq:p2}
\suff{0[01]^k}_{2p-2}=\suff{11[01]^*0[01]^k}_{2p-2}\cup
\suff{0[01]^*0[01]^k}_{2p-2}\cup 1[01]^{p-k-2}0[01]^k.
\end{equation}
For future reference, note that 
\[\{1[01]^{p-k-2}0[01]^k\,|\,0\leq k\leq p-3\}=\{[10]^k [01]^{p-k-1}
\,|\,2\leq k\leq p-1\}.\] For $k=p-2$ we simply have
$\suff{0[01]^{p-2}}_{2p-2}=\{00[01]^{p-2}, 10[01]^{p-2}\}$. Taking the
union of that last identity with the identities \eqref{eq:p2}, for all
$k=0,\dots,p-3$, we arrive at the formula
\begin{multline}\label{eq:p3}
\suff{0[01]^*}_{2p-2}= \suff{11[01]^*0[01]^*}_{2p-2}
\cup\suff{0[01]^*0[01]^*}_{2p-2}\cup \\\cup\{[10]^k[01]^{p-k-1}\,|\,1\leq k\leq p-1\},
\end{multline}
where the element $00[01]^{p-2}$ went into the second term and the
element $10[01]^{p-2}$ went into the third term on the right hand
side. We now substitute \eqref{eq:p3} into \eqref{eq:p1b} to get the
identity
\begin{multline*}
2^{[2p-2]}= \suff{11[01]^*}_{2p-2}\cup\suff{11[01]^*0[01]^*}_{2p-2}
\cup\suff{0[01]^*0[01]^*}_{2p-2}\cup \\\cup\{[10]^k[01]^{p-k-1}\,|\,0\leq k\leq p-1\}.
\end{multline*}
Combining this with the suffix $01$ we get
\begin{multline}\label{eq:p4}
\suff{01}_{2p}=\suff{11[01]^*01}_{2p}\cup\suff{11[01]^*0[01]^*01}_{2p}\cup
\suff{0[01]^*0[01]^*01}_{2p}\cup\\\cup\{[10]^k[01]^{p-k}\,|\,1\leq k\leq p-1\}.
\end{multline}
Substituting this into \eqref{eq:p1} we arrive at the formula
\begin{multline}\label{eq:p5}
\suff{1[10]^*}_{2p}=\suff{11}_{2p}\cup\suff{1[10]^*10}_{2p}\cup
\suff{11[01]^*01}_{2p}\cup\suff{11[01]^*0[01]^*01}_{2p}\cup\\\cup
\suff{0[01]^*0[01]^*01}_{2p}\cup\{[10]^k[01]^{p-k}\,|\,0\leq k\leq p-1\}.
\end{multline}
We now note following identities: $\suff{11}_{2p}\cup\suff{11[01]^*01}_{2p}=
\suff{11[01]^*}_{2p}$,
\begin{multline*}
\suff{1[10]^*10}_{2p}\cup\suff{11[01]^*0[01]^*01}_{2p}=
\suff{1[10]^*10}_{2p}\cup\suff{1[10]^*10[01]^*01}_{2p}=\\
=\suff{1[10]^*10[01]^*}_{2p},
\end{multline*}
and $\suff{0[01]^*0[01]^*01}_{2p}=\suff{00[10]^*01[01]^*}_{2p}$.
Substituting these back into~\eqref{eq:p5} will
yield~\eqref{eq:110even}.  

No new ideas are needed to show \eqref{eq:110odd}. All we have to do
is the retrace the argument used to show \eqref{eq:110even}.
Throughout the argument, all $\suff{}_{2p}$ and $\suff{}_{2p-2}$
should be replaced with $\suff{}_{2p+1}$ and $\suff{}_{2p-1}$.
Then~\eqref{eq:p0} and~\eqref{eq:p1} remain the same, subject to the
subscript change we just mentioned, while~\eqref{eq:p1b} gets replaced
with
\begin{equation}\label{eq:p1bodd}
2^{[2p-1]}=\suff{11[01]^*}_{2p-1}\cup\suff{0[01]^*}_{2p-1}\cup 1[01]^{p-1}.
\end{equation}
The identity~\eqref{eq:p1c} becomes
\[
\suff{0[01]^*}_{2p-1}=\suff{0}_{2p-1}\cup\suff{001}_{2p-1}\cup\dots\cup
\suff{0[01]^{p-2}}_{2p-1}\cup\suff{0[01]^{p-1}}_{2p-1},
\]
and we get 
\[
\suff{0[01]^k}_{2p-1}=\suff{11[01]^*0[01]^k}_{2p-1}\cup
\suff{0[01]^*0[01]^k}_{2p-1}\cup [01]^{p-k-1}0[01]^k,
\]
for $0\leq k\leq p-2$, and $\suff{0[01]^{p-1}}_{2p-1}=\{0[01]^{p-1}\}$. 
The identity \eqref{eq:p3} becomes
\begin{multline}\label{eq:p3odd}
\suff{0[01]^*}_{2p-1}= \suff{11[01]^*0[01]^*}_{2p-1}
\cup \suff{0[01]^*0[01]^*}_{2p-1}\cup \\\cup
\{0[10]^k[01]^{p-k-1}\,|\,0\leq k\leq p-1\}.
\end{multline}
Substituting this into \eqref{eq:p1bodd} yields
\begin{multline*}
2^{[2p-1]}= \suff{11[01]^*}_{2p-1}\cup 1[01]^{p-1}\cup\suff{11[01]^*0[01]^*}_{2p-1}
\cup\suff{0[01]^*0[01]^*}_{2p-1}\cup \\\cup\{0[10]^k[01]^{p-k-1}\,|\,0\leq k\leq p-1\}.
\end{multline*}
and combining with the suffix $01$ as above, and then substituting it
into the analog of \eqref{eq:p1} gives the analog of the
formula~\eqref{eq:p5}, which now says the following
\begin{multline}\label{eq:p5odd}
\suff{1[10]^*}_{2p+1}=\suff{11}_{2p+1}\cup\suff{1[10]^*10}_{2p+1}\cup
\suff{11[01]^*01}_{2p+1}\cup1[01]^p\cup\\\cup\suff{11[01]^*0[01]^*01}_{2p+1}\cup
\suff{0[01]^*0[01]^*01}_{2p+1}\cup\{0[10]^k[01]^{p-k}\,|\,0\leq k\leq p-1\}.
\end{multline}
Repeating the same transformations as in the $2p$-case we
derive~\eqref{eq:110odd}.  \qed 

\vskip5pt

We have now used an explicit constructive argument to prove the
Proposition~\ref{prop:110}. We feel, it is well-suited for explaining
the inner mechanics of the formulae~\eqref{eq:110even}
and~\eqref{eq:110odd}. However, it is also possible to give a~much
shorter implicit argument. To save space, we restrict ourselves to
giving a~sketch of how to show~\eqref{eq:110even} in two
steps. Step~1: count the number of elements on both side and derive
that they are both equal to $\frac{2}{3}(4^p-1)$. Note that we do not
know yet that the sets on the right hand side are disjoint, so
overlaps are counted multiple times.  Step 2: show that every element
from the left hand side can be found on the right hand side. This can
be done by considering several different cases.  Once this is done we
know both that the two sides are equal and that the sets on the right
hand side are disjoint. The identity \eqref{eq:110odd} can be shown
exactly the same, with the number of elements on both sides being
equal to $\frac{1}{3}(4^p-1)$.

\subsection{The main bijection and the proof of our main theorem} $\,$

\nin We let $\Sigma\suf\alpha$ denote the set of all subsets from
a~family $\Sigma$ which end on $\alpha$, in other words
$\Sigma\suf\alpha:=\Sigma\cap\suf\alpha$. Clearly, all the
decompositions above can be intersected with $\Sigma$. This will give
a~number of different decompositions of $\Sigma$, such as
\[\Sigma=\Sigma\suf{0[01]^*}\cup\Sigma\suf{11[01]^*}\cup\{[01]^{n/2}\}\]

\begin{prop}\label{prop:Phi}
For any integer $n\geq 5$, and any $0\leq t\leq n-1$, there exists
a~bijection
\[\Phi_t^n:C^n_t\suf{0[01]^*}\ra C^n_{t+1}\suf{1[10]^*},\] 
such that for all $S\in\suf{0[01]^*}_n$, we have $S\subseteq\Phi_t^n(S)$.
\end{prop}

\pr Assume first that $n=2p$ is even, and compare the
formula~\eqref{eq:001even} with~\eqref{eq:110even}. We see a~strong
similarity, and define the bijection $\Phi_t^n$ by the rules
\[
\begin{array}{rcl}
\alpha 10[01]^k&\mapsto& \alpha 11[01]^k,\\
\alpha1[10]^m 00[01]^k &\mapsto &\alpha 1[10]^m 10[01]^k,\\
\alpha00[10]^m 00[01]^k &\mapsto &\alpha 00[10]^m 01[01]^k,\\
\,[10]^k 00[01]^{p-k-1}&\mapsto & [10]^k 01[01]^{p-k-1},
\end{array}
\]
for all $k,m\geq 0$, and all strings~$\alpha$. When $n=2p+1$ is odd,
we compare the formula~\eqref{eq:001odd} with~\eqref{eq:110odd}
instead, and the last rule of the bijection gets changed to
\[
\begin{array}{rcl}
0[10]^k 00[01]^{p-k-1}& \mapsto &0[10]^k 01[01]^{p-k-1},\\
0[01]^p& \mapsto &1[01]^p,
\end{array}
\]
for all~$k$.
\qed

\begin{df}\label{df:psi}
Assume $2\leq t\leq n$, and consider bijections 
\[
\begin{array}{rclcrcl}
\gamma:C_t^n\suf{11[01]^*} & \ra & C_{t-1}^{n-1}(1[10]^*) & \quad & 
\rho:C_{t-2}^{n-1}\suff{0[01]^*} & \ra & C_{t-2}^n\suff{00[10]^*} \\
\alpha 1[10]^k 1 & \mapsto & \alpha 1[10]^k & \quad &
\alpha 0[01]^*     & \mapsto & \alpha0[01]^*0
\end{array}
\]
where $\alpha$ is any string, and $k$ is any positive
integer.\footnote{Note how we use the facts that $11[01]^k=1[10]^k1$
  and $00[10]^k=0[01]^k0$.} We then define a~bijection
\[\Psi^n_t:C_t^n\suf{11[01]^*}\ra C_{t-2}^n\suf{00[10]^*}\]
as a~composition
$\Psi^n_t:=\rho\circ(\Phi_{t-2}^{n-1})^{-1}\circ\gamma$.
\end{df}

Let $M_r^n=\{S\subseteq[n]\,|\, |S|\equiv r\mod 3\}$.

\begin{prop}
For all $t\geq 1$, there exists a bijection 
\[\Lambda:M_0^{6t}\sm[01]^{3t}\ra M_1^{6t},\]
such that either $S\subset\Lambda(S)$ or $\Lambda(S)\subset S$.  Under
this bijection we have $\Lambda([1]^{6t})=[1]^{6t-2}00$.
\end{prop}

\pr
By Corollary~\ref{crl:exp} we have 
\[M_0^{6t}\sm[01]^{3t}=\bigcup_{k=0}^{2t-1}C_{3k}^n\suff{0[01]^*}\cup
\bigcup_{k=1}^{2t}C_{3k}^n\suff{11[01]^*},\]
\[M_1^{6t}=\bigcup_{k=0}^{2t-1}C_{3k+1}^n\suff{1[10]^*}\cup
\bigcup_{k=1}^{2t}C_{3k-2}^n\suff{00[10]^*},\] where all the unions
are disjoint. We now define $\Lambda$ by saying that the restriction
of $\Lambda$ to $C_{3k}^n\suff{0[01]^*}$ is equal to $\Phi_{3k}^n$,
and the restriction of $\Lambda$ to $C_{3k}^n\suff{11[01]^*}$ is equal
to $\Psi_{3k}^n$. By what is proved until now, this is clearly
a~bijection. \qed

\vskip5pt

\nin We are now ready to prove our main theorem.

\vskip5pt

\nin{\bf Proof of Theorem~\ref{thm:6n}.}  We have a matching where the
only unmatched sets are $[01]^{3t}$ and $[1]^{6t-2}00$. We fix that by
using an augmenting path
\[[01]^{3t}\rightsquigarrow [01]^{3t-1}11\ra[01]^{3t-1}10\rightsquigarrow
[01]^{3t-2}1110\ra[01]^{3t-2}1100\rightsquigarrow [1]^{6t-2}00,\] 
where the edges $[01]^{3t-1}11\ra[01]^{3t-1}10$
and $[01]^{3t-2}1110\ra[01]^{3t-2}1100$ are matching edges. \qed

\begin{dcc}
Theorem~\ref{thm:6n} means that whenever the number of processes is
divisible by $6$, the Weak Symmetry Breaking task can be solved in 3
rounds. In particular, there are infinitely many values for the number
of processes, for which this task can be solved using a~constant
number of rounds.
\end{dcc}

The smallest values of $n$ which are not covered by
Theorem~\ref{thm:6n} are $n=10,14,15$. The binomial Diophantine
equations associated to $n=10$ and to $n=14$ do not have primitive
solutions. For $n=15$ we do find several primitive solutions, for
example $x_1=x_3=x_5=x_{10}=1$, $x_4=x_6=x_{13}=-1$, and for all other
$i$ we take $x_i=0$. This corresponds to the following arithmetic
identity:
\[\binom{15}{1}+\binom{15}{3}+\binom{15}{5}+\binom{15}{10}=
\binom{15}{0}+\binom{15}{4}+\binom{15}{6}+\binom{15}{13}=6476.\] 
A computer search can then be used to find a~comparable
matching between disjoint cardinal proper families of subsets of
$[15]$, implying $\msb(15)\leq 3$.

\begin{prop} 
The binomial Diophantine equation associated to $n$ has solutions if
and only if $n$ is not a~prime power. Furthermore, there are
infinitely many values of $n$, say $n=6t$, for arbitrary natural
number $m$, for which the binomial Diophantine equation associated to
$n$ has a~primitive solution. 
\end{prop}
\pr If $n=p^m$, then all the binomial coefficients
$\binom{p}{1},\dots,\binom{p}{p-1}$ are divisible by $p$, so obviously
the binomial Diophantine equation associated to $n$ has no
solutions. Otherwise, the greatest common divisor of these
coefficients is equal to $1$, and so a~solution can be found by
Euclidean algorithm.

For $n=6t$ we have an identity
\begin{equation}\label{eq:6t}
\binom{6t}{0}+\binom{6t}{3}+\binom{6t}{6}+\dots+\binom{6t}{6t-3}=
\binom{6t}{1}+\binom{6t}{4}+\binom{6t}{7}+\dots+\binom{6t}{6t-2},
\end{equation} 
so the following is a primitive solution: $x_1=x_4=\dots=x_{6t-2}=1$,
$x_3=x_6=\dots=x_{6t-3}=-1$, and all other coefficients are equal
to~$0$.  \qed

\subsection{Example $t=1$} $\,$

\nin When $t=1$ we are dealing with the subsets of the set $[6]$. We
have $|M_0^6\sm[01]^3|=|M_1^6|=21$ and we need to match the elements
of these two sets with each other. To start with we have
$C_0^6\suff{0[01]^*}=000000$, $C_1^6\suff{1[10]^*}=000001$,
furthermore $000000\in C_0^6\suff{00[10]^*00[01]^*}$, and hence
$\Phi^6_0(000000)=000001$. Similarly, $C_6^6\suff{11[01]^*}=111111$,
$C_4^6\suff{00[10]^*}=111100$, and $\Psi_6^6(111111)=111100$. It
remains to mutually match the $14$-element sets $C_3^6\suff{0[01]^*}$
and $C_4^6\suff{1[10]^*}$, and the $5$-element sets
$C_3^6\suff{11[01]^*}$ and $C_1^6\suff{00[10]^*}$.

We start with the two $14$-element sets. In this case the
formula~\eqref{eq:001even} simplifies to
\[C_3^6\suff{0[01]^*}=C_3^6\suff{10[01]^*}\cup C_3^6\suff{1[10]^*00[01]^*},\]
where the first set in the union has $9$ elements, and the second one
has $5$ elements.  Similarly, the formula~\eqref{eq:110even}
simplifies to
\[C_4^6\suff{1[10]^*}=C_4^6\suff{11[01]^*}\cup C_4^6\suff{1[10]^*10[01]^*}.\]
Our matching rule is now the following
\[
\begin{array}{ccccccc}
C_3^6\suff{1{\bf 0}[01]^*}& \! &C_4^6\suff{1{\bf 1}[01]^*}&\,&
C_3^6\suff{1[10]^*{\bf 0}0[01]^*}&\! &C_4^6\suff{1[10]^*{\bf 1}0[01]^*}\\
1{\bf 0}0101 & \mapsto & 1{\bf 1}0101&\quad&1101{\bf 0}0&\mapsto&1101{\bf 1}0\\
011{\bf 0}01 & \mapsto & 011{\bf 1}01&\quad&0111{\bf 0}0&\mapsto&0111{\bf 1}0\\
101{\bf 0}01 & \mapsto & 101{\bf 1}01&\quad&1011{\bf 0}0&\mapsto&1011{\bf 1}0\\
00111{\bf 0} & \mapsto & 00111{\bf 1}&\quad&1110{\bf 0}0&\mapsto&1110{\bf 1}0\\
01011{\bf 0} & \mapsto & 01011{\bf 1}&\quad&11{\bf 0}001&\mapsto&11{\bf 1}001\\
10011{\bf 0} & \mapsto & 10011{\bf 1}&&&&\\
01101{\bf 0} & \mapsto & 01101{\bf 1}&&&&\\
10101{\bf 0} & \mapsto & 10101{\bf 1}&&&&\\
11001{\bf 0} & \mapsto & 11001{\bf 1}&&&&
\end{array}
\]

For the above-mentioned $5$-element sets we need to use the bijection
$\Psi_3^6$, whose definition is somewhat more complicated. The
composition from Definition~\ref{df:psi} yields in our case the following maps:
\[
\begin{array}{ccccccc}
100011&\longrightarrow&1000{\bf 1}&\longrightarrow&1000{\bf 0}&\longrightarrow&100000\\
010011&\longrightarrow&0100{\bf 1}&\longrightarrow&0100{\bf 0}&\longrightarrow&010000\\
001101&\longrightarrow&001{\bf 1}0&\longrightarrow&001{\bf 0}0&\longrightarrow&001000\\
000111&\longrightarrow&0001{\bf 1}&\longrightarrow&0001{\bf 0}&\longrightarrow&000100\\
001011&\longrightarrow&00{\bf 1}01&\longrightarrow&00{\bf 0}01&\longrightarrow&000010
\end{array}
\]

Finally, we need to alter our matching one time since $010101$ and
$111100$ are not matched, alternatively, we can think that $111100$ is
matched to $111111$, which needs the same modification. This is done
as is described in the proof of Theorem~\ref{thm:6n}. We break down
the bonds $010110\mapsto 010111$ and $011100\mapsto 011110$, and take
the following 3 edges as the new matching edges: $010101\mapsto 010111$,
$010110\mapsto 011110$, and $011100\mapsto 111100$.


\section{Current bounds for the symmetry breaking number} \label{sect:11}

We strongly believe that the techniques developed in this paper can be
extended to deal with many other values of~$n$. This has recently been
confirmed as follows.

\begin{df} \cite[Definitions~1.1, 1.2]{bid}. $\,$
 Assume that $n$ is a natural number and that for some numbers $0\leq
a_1<\dots<a_k\leq n$ and $0\leq b_1<\dots<b_m\leq n$, we have an
equality
\begin{equation} \label{eq:bin}
\binom{n}{a_1}+\dots+\binom{n}{a_k}=\binom{n}{b_1}+\dots+\binom{n}{b_m}.
\end{equation}
We call such an identity a~{\bf binomial identity}.

Let $\Sigma$, resp.\ $\Lambda$, be set of all subsets of $[n]$, with
cardinalities $a_1,\dots,a_k$, resp.\ $b_1,\dots,b_m$. We say that the
binomial identity \eqref{eq:bin} is {\bf orderable} if there exists
a~bijection $\Phi:\Sigma\ra\Lambda$, such that for all $S\in\Sigma$ we
either have $S\subseteq\Phi(S)$ or $S\supseteq\Phi(S)$.
\end{df}

In this paper we have constructed a complicated explicit bijection for
the case $n=6t$, $k=m=2t$, $\{a_1,\dots,a_{2t}\}=\{0,3,\dots,6t-3\}$,
and $\{b_1,\dots,b_{2t}\}=\{1,4,\dots,6t-2\}$, see
Section~\ref{sect:10}. It would be interesting to see whether
a~simpler bijection can be found.

Recently, we proved the following combinatorial theorem.

\begin{thm}\label{conj:main}  \cite[Theorem~1.3]{bid}. 
All binomial identities are orderable.
\end{thm}

It has been shown in \cite{bid} that, together with
Theorem~\ref{thm:main}, this implies the following bound on the
symmetry breaking number.

\begin{thm}\cite[Theorem~3.2]{bid}. \label{conj:prim}
Assume the binomial Diophantine equation associated to $n$ has
a~primitive solution, then we have $\msb(n)\leq 3$.
\end{thm}

\vskip5pt

Our current knowledge about $\msb(n)$ is summarized in
the Table~\ref{table:sb}.

\begin{table}[hbt] 
\[\begin{array}{l|l}
\text{Bound} & \text{Source}\\ \hline
 \msb(n)=\infty \text{ if and only if } n \text{ is a prime power } & 
\text{\cite{CR0,CR1,CR2}}\\ [0.2cm]
 \msb(n)=O(n^{q+3}),
 \text{ if } n \text{ is not a~prime power and } & \text{\cite{ACHP}}\\
q \text{ is the largest prime power in
the prime factorization of } n &\\  [0.2cm]
 \msb(n)\geq 2& \text{\cite{wsb6}} \\  [0.2cm]
 \msb(6t)\leq 3,\text{ for all }t\geq 1& \text{Theorem~\ref{thm:6n} above;} \\
&\text{the case $t=1$ in \cite{wsb6}}\\  [0.2cm]
 \msb(n)\leq 3,\text{ if the binomial Diophantine equation associated } & \text{\cite{bid}}\\
\text{to $n$ has
a~primitive solution}&
\end{array}\]
\caption{The known bounds of $\msb(n)$.}
\label{table:sb}
\end{table}

In general, we feel that the work presented in this paper is
suggesting that we need to change our paradigm. When looking for lower
bounds for the complexity of the distributed protocols solving Weak
Symmetry Breaking, the focus needs to shift from the number of
processes $n$ itself to the sizes of the coefficients in the solution
of the binomial Diophantine equation associated to~$n$.

\vspace{10pt}

\nin{\bf Acknowledgments.} I would like to thank Maurice Herlihy and
Sergio Rajsbaum for discussions and encouragement for writing up these
results. 




\section{Appendix: Path building kit}

\nin In this appendix we list different edge paths in $\gn$ which are
used elsewhere in the paper. These paths are alternating with respect
to some give matching $\mu$, and we use $\stackrel\mu\longrightarrow$
to denote edges belonging to the matching, while $\rightsquigarrow$
denotes all other edges.

\begin{figure}[hbt]
\begin{tikzcd}
 n \s x_2\s\dots\s x_k\s x_{k+1}\s\dots \arrow{d}{\mu} & 
 n , x_2\s  \dots \s x_k\s x_{k+1}\s\dots \arrow{d}{\mu} \\
 n, x_2\s  \dots \s  x_k \s x_{k+1}\s\dots\arrow[squiggly]{d} & 
 n\s x_2\s  \dots \s x_k \s x_{k+1}\s\dots \arrow[squiggly]{d} \\
 n, x_2\s  \dots \s  x_k,x_{k+1}\s\dots \arrow{d}{\mu} & 
 n\s x_2\s  \dots \s x_k,x_{k+1}\s\dots \arrow{d}{\mu} \\
 n\s x_2\s  \dots \s  x_k,x_{k+1}\s\dots \arrow[squiggly]{d} & 
 n, x_2\s  \dots \s  x_k,x_{k+1}\s\dots \arrow[squiggly]{d} \\
 n\s x_2\s  \dots \s  x_{k+1}\s x_{k}\s\dots & 
 n, x_2\s  \dots \s  x_{k+1}\s x_{k}\s\dots
\end{tikzcd}
\caption{Paths $\swap^I_k$ and $\swap^{II}_k$, for $3\leq k\leq n-1$.}
\label{table:swapk}
\end{figure}
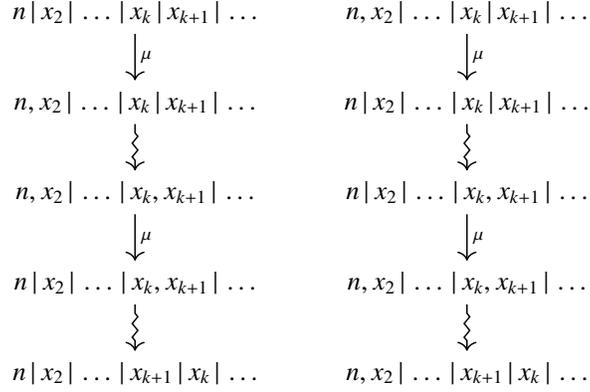

\begin{figure}[hbt]
\begin{tikzcd}
n \s x_2\s x_3\s\dots \s x_{n-1}\s x_{n} \arrow{d}{\mu} &
n,x_2\s x_3\s   \dots \s x_{n-1}\s x_{n} \arrow{d}{\mu}\\
n,x_2\s x_3\s   \dots \s x_{n-1}\s x_{n} \arrow[squiggly]{d} &
n\s x_2\s x_3\s \dots \s x_{n-1}\s x_{n} \arrow[squiggly]{d}\\
n, x_2\s x_3\s  \dots \s x_{n-1},  x_{n} \arrow{d}{\mu} &
n\s x_2, x_3\s  \dots \s x_{n-1}\s x_{n} \arrow{d}{\mu}\\
n\s x_2\s x_3\s \dots \s x_{n-1},  x_{n} \arrow[squiggly]{d} & 
n, x_2,  x_3\s  \dots \s x_{n-1}\s x_{n} \arrow[squiggly]{d}\\
n\s x_2, x_3\s  \dots \s x_{n-1},  x_{n} \arrow{d}{\mu} &
n, x_2,  x_3\s  \dots \s x_{n-1}, x_{n} \arrow{d}{\mu}\\
n, x_2,  x_3\s  \dots \s x_{n-1}, x_{n} \arrow[squiggly]{d} & 
n\s x_2, x_3\s  \dots \s x_{n-1}, x_{n} \arrow[squiggly]{d}\\
n,  x_2, x_3\s  \dots \s x_{n-1}\s x_{n} \arrow{d}{\mu} & 
n\s x_3 \s x_2\s\dots \s x_{n-1}, x_{n} \arrow{d}{\mu}\\
n\s x_2, x_3\s  \dots \s x_{n-1}\s x_{n} \arrow[squiggly]{d} &
n,x_3\s x_2\s   \dots \s x_{n-1}, x_{n} \arrow[squiggly]{d}\\
n\s x_3\s x_2\s \dots \s x_{n-1}\s x_{n} &
n,x_3\s x_2\s   \dots \s x_{n-1}\s x_{n}
\end{tikzcd}
\caption{Paths $\swap^I_2$ and $\swap^{II}_2$.}
\label{table:swap2}
\end{figure}

\begin{figure}[hbt]
\begin{tikzcd}
 n \s x_2\s x_3\s\dots\arrow{d}{\mu} & 
 x_1\s \dots \s x_{k-1}\s n\s x_{k+1}\s\dots \arrow{d}{\mu}\\
 n, x_2\s x_3\s\dots\arrow[squiggly]{d} & 
 x_1\s \dots \s x_{k-1}\s n,x_{k+1}\s\dots \arrow[squiggly]{d} \\
 x_2\s n\s x_3\s \dots &
 x_1\s \dots \s x_{k-1}\s x_{k+1}\s n\s\dots  
\end{tikzcd}
\caption{On the left hand side we have the alternating path $\up^I_1$,
  which is legal if either $x_2\notin V$ or if $x_2\in\Omega$. On the
  right hand side we have alternating path $\up^I_k$, for $2\leq k\leq
  n-1$, which is legal if either $x_1\notin V$, or if $x_1\s\dots\s
  x_{k-1}\in\Omega$ and $x_1\s\dots\s x_{k-1}\s x_{k+1}\in\Omega$.  }
\label{table:up1}
\end{figure}
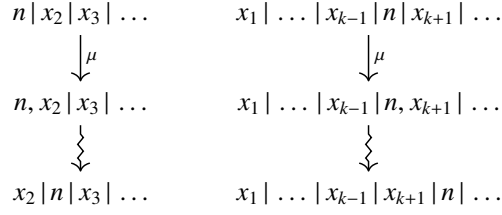

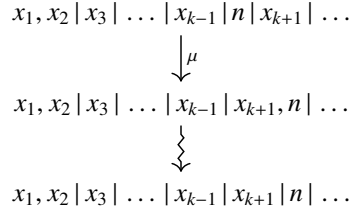
\begin{figure}[hbt]
\begin{tikzcd} 
 x_1,x_2\s x_3\s \dots \s x_{k-1}\s n\s x_{k+1}\s\dots \arrow{d}{\mu}\\
 x_1,x_2\s x_3\s \dots \s x_{k-1}\s x_{k+1},n  \s\dots \arrow[squiggly]{d}\\
 x_1,x_2\s x_3\s \dots \s x_{k-1}\s x_{k+1}\s n\s\dots
\end{tikzcd}
\caption{Path $\up^{II}_k$, for $3\leq k\leq n-1$: legal if either
  $\{x_1,x_2\}\not\subseteq V$, or if $\{x_1,x_2\}\s x_3\s\dots\s
  x_{k-1} \in\Omega$ and $\{x_1,x_2\}\s x_3\s\dots\s x_{k-1}\s
  x_{k+1}\in\Omega$.}
\label{table:up2k}
\end{figure}

\begin{figure}[hbt]
\begin{tikzcd}
n,a_2  \s n-1\s a_4\s\dots\arrow{d}{\mu} &
n-1,n\s a_3\s a_4\s\dots \arrow{d}{\mu} \\
n\s a_2\s n-1\s a_4\s\dots\arrow[squiggly]{d}&
n\s n-1\s a_3\s a_4\s\dots\arrow[squiggly]{d} \\
n\s a_2,n-1  \s a_4\s \dots\arrow{d}{\mu}&
n\s n-1\s a_3, a_4\s\dots\arrow{d}{\mu}   \\
n,a_2,n-1    \s a_4\s \dots\arrow[squiggly]{d}&
n-1,n\s a_3, a_4\s\dots\arrow[squiggly]{d} \\
n-1\s n,a_2  \s a_4\s \dots\arrow{d}{\mu}&
n-1\s n\s a_3, a_4\s\dots\arrow{d}{\mu}   \\
n-1\s n\s a_2\s a_4\s \dots\arrow[squiggly]{d}&
n-1\s n, a_3, a_4\s\dots \arrow[squiggly]{d} \\
n-1\s n\s   a_2,a_4\s \dots\arrow{d}{\mu} &
n-1\s a_3\s n,a_4\s\dots\arrow{d}{\mu}  \\
n-1\s   n,a_2,a_4  \s \dots\arrow[squiggly]{d} &
n-1\s a_3\s n\s a_4\s\dots\arrow[squiggly]{d} \\
n-1\s a_2\s   n,a_4\s \dots\arrow{d}{\mu}&
n-1, a_3\s n\s a_4\s\dots  \\
n-1\s a_2\s n\s a_4\s \dots\arrow[squiggly]{d} \\
n-1,a_2\s   n\s a_4\s \dots
\end{tikzcd}
\caption{Paths $\specup^{II}$ and $\up^{II}_2$.}
\label{table:ups}
\end{figure}

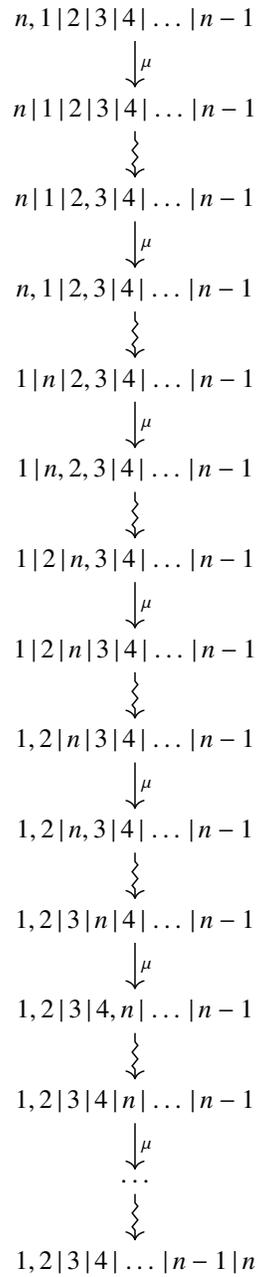
\begin{figure}[hbt]
\begin{tikzcd}
 n,1\s 2 \s 3\s 4 \s\dots\s n-1   \arrow{d}{\mu} \\
 n\s 1\s 2\s 3\s 4\s\dots\s n-1 \arrow[squiggly]{d}\\
 n\s 1\s 2,3\s 4  \s\dots\s n-1   \arrow{d}{\mu} \\
 n,1 \s  2,3\s 4  \s\dots\s n-1 \arrow[squiggly]{d}\\
 1\s n\s 2,3\s 4  \s\dots\s n-1   \arrow{d}{\mu} \\
 1\s n,2,3 \s 4   \s\dots\s n-1 \arrow[squiggly]{d}\\
 1\s 2\s n,3\s 4  \s\dots\s n-1   \arrow{d}{\mu} \\
 1\s 2\s n\s 3\s 4\s\dots\s n-1 \arrow[squiggly]{d}\\
 1,2  \s n\s 3\s 4\s\dots\s n-1   \arrow{d}{\mu} \\
 1,2\s n,3 \s 4   \s\dots\s n-1 \arrow[squiggly]{d}\\
 1,2\s 3\s n \s 4 \s\dots\s n-1   \arrow{d}{\mu} \\
 1,2\s 3\s 4, n\s\dots\s n-1 \arrow[squiggly]{d}\\
 1,2\s 3\s 4\s n\s\dots\s n-1   \arrow{d}{\mu} \\
\dots\arrow[squiggly]{d}\\
 1,2\s 3\s 4\s \dots \s n-1\s n
\end{tikzcd}
\caption{The final part for the path in Lemma~\ref{lm:sc-}.}
\label{table:star}
\end{figure}

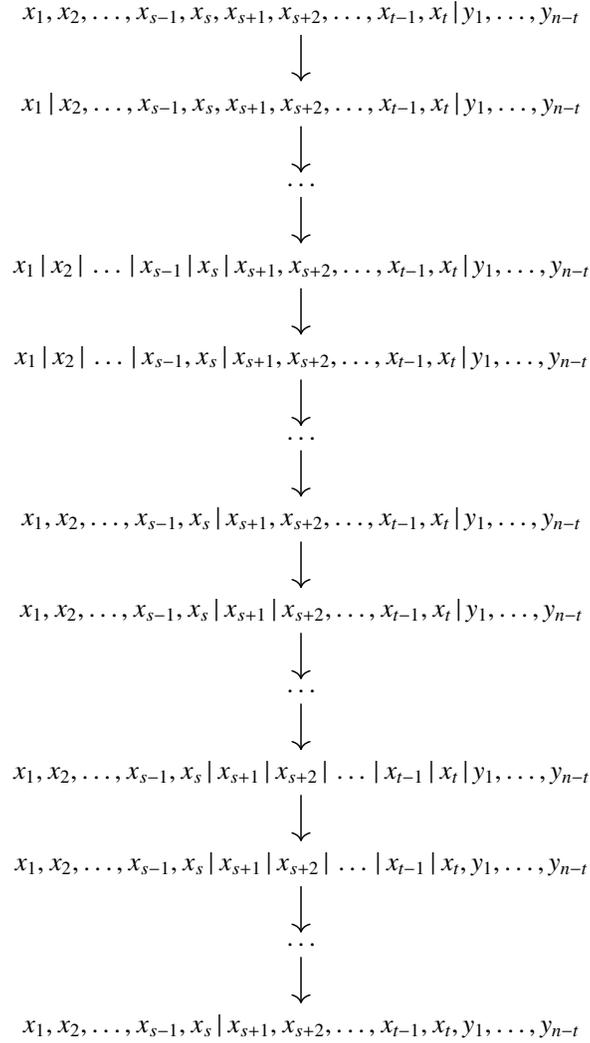
\begin{figure}[hbt]
\begin{tikzcd}
 x_1,x_2,\dots,x_{s-1},x_s,x_{s+1},x_{s+2},\dots,x_{t-1},x_t \s y_1,\dots,y_{n-t} \arrow{d} \\
 x_1\s x_2,\dots,x_{s-1},x_s,x_{s+1},x_{s+2},\dots,x_{t-1},x_t\s y_1,\dots,y_{n-t}\arrow{d} \\
\dots\arrow{d} \\
 x_1\s x_2\s \dots \s x_{s-1}\s x_s\s x_{s+1},x_{s+2},\dots,x_{t-1},x_t \s y_1,\dots,y_{n-t} \arrow{d} \\
 x_1\s x_2\s \dots \s x_{s-1},x_s\s x_{s+1},x_{s+2},\dots,x_{t-1},x_t \s y_1,\dots,y_{n-t} \arrow{d} \\
\dots\arrow{d} \\
 x_1,x_2,\dots,x_{s-1},x_s\s x_{s+1},x_{s+2},\dots,x_{t-1},x_t \s y_1,\dots,y_{n-t} \arrow{d} \\
 x_1,x_2,\dots,x_{s-1},x_s\s x_{s+1}\s x_{s+2},\dots,x_{t-1},x_t \s y_1,\dots,y_{n-t} \arrow{d} \\
\dots\arrow{d} \\
 x_1,x_2,\dots,x_{s-1},x_s\s x_{s+1}\s x_{s+2}\s\dots\s x_{t-1}\s x_t \s y_1,\dots,y_{n-t} \arrow{d} \\
 x_1,x_2,\dots,x_{s-1},x_s\s x_{s+1}\s x_{s+2}\s\dots\s x_{t-1}\s x_t, y_1,\dots,y_{n-t} \arrow{d} \\
\dots\arrow{d} \\
 x_1,x_2,\dots,x_{s-1},x_s\s x_{s+1}, x_{s+2},\dots,x_{t-1},x_t, y_1,\dots,y_{n-t}  
\end{tikzcd}
\caption{The standard path $p_{S,T}$ for $S=(x_1,\dots,x_s)$, $T=(x_1,\dots,x_t)$.}
\label{fig:pst1}
\end{figure}

\end{document}